%% file: main.tex
\documentclass[conference]{IEEEtran}
\IEEEoverridecommandlockouts

\input{datalabmacros}

\input{individualmacros}

\usepackage{cite}
\usepackage{amsmath,amssymb,amsfonts}
\usepackage{algorithmic}

\usepackage{graphicx}
\usepackage{textcomp}
\usepackage{xcolor}
\def\BibTeX{{\rm B\kern-.05em{\sc i\kern-.025em b}\kern-.08em
    T\kern-.1667em\lower.7ex\hbox{E}\kern-.125emX}}
\begin{document}

\title{Synthesizing Scoring Functions for Rankings Using Symbolic Gradient Descent
}

\author{
\IEEEauthorblockN{Zixuan Chen}
\IEEEauthorblockA{\textit{Northeastern University} \\
Boston, USA \\ chen.zixu@northeastern.edu \\
\orcidlink{0000-0003-2872-1865} 0000-0003-2872-1865}
\and
\IEEEauthorblockN{Panagiotis Manolios}
\IEEEauthorblockA{\textit{Northeastern University} \\
Boston, USA \\ p.manolios@northeastern.edu \\
\orcidlink{0000-0003-0519-9699} 0000-0003-0519-9699}
\and
\IEEEauthorblockN{Mirek Riedewald}
\IEEEauthorblockA{\textit{Northeastern University} \\
Boston, USA \\ m.riedewald@northeastern.edu \\
\orcidlink{0000-0002-6102-7472} 0000-0002-6102-7472}
}

\maketitle

\begin{abstract}
Given a relation and a ranking of its tuples, but no information about
the ranking function, we are interested in synthesizing
simple scoring
functions that reproduce the ranking. Our system \RankHow
identifies linear scoring functions that minimize position-based
error, while supporting flexible constraints on their weights.
It is based on a new formulation
as a mixed-integer linear program (MILP). While MILP is NP-hard
in general, we show that \RankHow is orders of magnitude faster than
a tree-based algorithm that guarantees polynomial time
complexity (PTIME) in the number of input tuples by reducing
the MILP problem to many linear programs (LPs).
We hypothesize that this is caused by 2 properties:
First, the PTIME algorithm is equivalent to a naive evaluation strategy
for the MILP program. Second, MILP solvers rely on advanced heuristics
to reason holistically about the entire program, while the PTIME algorithm
solves many sub-problems in isolation.

To further improve \RankHow's scalability, we propose a novel
approximation technique called symbolic gradient descent (\SymGD).
It exploits problem structure to more quickly find local minima of the error function.
Experiments demonstrate that \RankHow can solve realistic problems,
finding more accurate linear scoring functions than the state of the art.
\end{abstract}


\section{Introduction}
\label{sec:introduction}
Rankings play an essential role, from personal life choices to business decisions:
they are prevalent in sports, where journalists and fans attempt to rank anything
and anyone from teams to coaches and individual players.
They help shoppers decide which product to purchase and let travelers more quickly
identify the right hotel or flight ticket.
Top-ranked universities attract applications from the best students in a matching
process driven by university rankings \cite{csrankings, usnews}.
And there exists a rich history of algorithms for problems like $k$-shortest paths \cite{1971-managementscience-yen, 1998-siamjc-eppstein, 1999-wae-jimenez},
top-$k$ queries \cite{2008-acmcs-ilyas, 2019-sigmodrec-rahul},
and ranked enumeration or \emph{any}-$k$ queries \cite{2021-icdt-deep, 2021-icdt-bourhis, 2020-pvldb-tziavelis, 2021-pvldb-tziavelis, 2022-icde-tziavelis}.

When a ranking is published without a ranking function, or when the process
of obtaining a ranking is opaque, it raises questions around algorithmic
decision making and fairness. To this end, we are interested in
\textbf{efficiently synthesizing simple linear scoring functions
that approximate it well}.

\begin{example}[NBA MVP selection]
\label{ex:nba-mvp}
Every year, the NBA's Most Valuable Player (MVP) is selected by a panel of
sportswriters and broadcasters from the US and Canada.
Each panelist chooses their top-5 players, assigning them 10, 7, 5, 3, and 1 points,
respectively. The final ranking, and thus the MVP, is based on point totals.
This protocol does not reveal how player features impacted the process, and a
panelist may not be able to quantitatively derive why they placed one player
above another.

For each player, a wealth of data exists, including feature attributes such as
points scored ($P$), rebounds ($R$), assists ($A$), steals ($S$), and
blocks ($B$). Using these five attributes, our system \RankHow determines
that when sorting by scoring function $0.02R + 0.14A + 0.84B$,
the panelists' top-3 ranking among all players who received at least one vote can
be recovered exactly.

\RankHow supports powerful \emph{constraints} for exploring alternative
scoring functions. These can enforce prior knowledge and ``realism''.
For example, points scored should feature prominently in NBA-player ranking.
After constraining the coefficient of $P$ to
be at least 0.1, \RankHow discovers that
scoring function $0.1P + 0.14R + 0.17A + 0.1S + 0.49B$
would also recover the panel's top-3 ranking.
One could similarly enforce upper or lower bounds on the sum of
selected coefficients, e.g., on all defensive skills.
Many other constraint types can be easily integrated into \RankHow:
No top-10 player should be placed more than 2 positions higher
or lower compared to the given ranking. The number-1 player,
Nikola Joki\'c, must be in position 1. For all players ranked 1 to 100,
a player ranked $i$-th in the original ranking must be ranked in
range $\lfloor 0.9 i \rfloor$ to $\lceil 1.1 i \rceil$.
Nikola Joki\'c must be ranked higher than Jayson Tatum.
And so on.
\end{example}

Our goal is \emph{not} to find the most accurate
scoring function, but the most accurate \emph{simple} scoring function.
Constructing a perfect scoring function is straightforward:
Given the original ranking and all data tuples,
we can simply use the data table directly! For instance,
if tuple $(3,2)$ is ranked at position 1 and tuple $(4,1)$ at position 2,
the corresponding function is:
``if $A_1=3$ and $A_2=2$ then return 1;
else if $A_1=4$ and $A_2=1$ then return 2.''
It is similar to a sufficiently grown decision tree where each tuple
appears in its own leaf node. This approach works here, because,
in contrast to learning-to-rank, there is no need to generalize to
unseen inputs.

Our approach follows the spirit of
Rudin~\cite{19-natmi-rudin} and of the famous Occam's razor,
which intuitively states that one should pursue
the simplest explanation possible for an observed phenomenon.
The arguably simplest possible scoring function is a linear combination of
the given original attributes. Finding such a function is meaningful,
because in the context of ranking, it is very natural to list desirable
(and undesirable) properties and then determine
how much they matter in the big picture of making a decision. E.g., should one
pay more attention to points scored vs assists vs rebounds?
(For undesirable properties, e.g., turnovers caused, the column is simply
converted to negative values.) Furthermore,
\emph{linear scalarization} represents one of the most popular approaches
for multi-objective optimization.
And linear scoring functions are arguably by far the most widely used
type of scoring function in the data management community
\cite{01-sigmod-hristidis-prefer,06-vldb-xin-indexing-ranked-queries,10-fskd-wang,10-icde-vlachou,10-vldb-vlachou,11-tkde-vlachou,12-icde-he,13-dasfaa-chester,13-sigmod-vlachou,14-dasfaa-jin,14-tkde-he,14-vldb-zhang,15-icde-chen,15-pvldb-gao,16-tkde-xu,18-sigmod-yang,18-vldb-asudeh-stable-rankings,19-sigmod-asudeh-fair-ranking,21-sigmod-wang,23-icde-wang,23-acmcs-zehlike-fair-ranking-suvey-part-1,23-pvldb-chen,24-icde-wang,19-pvldb-shetiya,22-icde-xiao}.

\textbf{How to use \RankHow.}
After applying \RankHow to a given ranking and dataset $R$, there are
two possible outcomes: the ranking produced by the simple linear scoring function
is ``good enough'', or it is not. If the former, then the user can continue to search
for alternative linear functions by providing additional constraints
as illustrated in \Cref{ex:nba-mvp}. 
Otherwise, i.e., if even the optimal simple ranking function's error
is too high, then one can use intuition, prompt engineering,
or other techniques to discover and add derived attributes to the input data
until a ``good enough'' ranking is obtained.
For instance, from the number of points scored and the time an NBA player
spent on the court, we can derive a new attribute points-per-minute.
Now \RankHow can synthesize a linear function over the increased
attribute space, but this function is actually non-linear in the original space.
Generalizing well-known kernels, e.g., polynomial or RBF kernels for SVMs,
we can use any function over the original attributes $A_1,\ldots, A_m$,
including those containing logarithms, polynomials, exponents, and so on.
In this paper we mostly focus on only using the original attributes
to determine how far even the simplest functions can get us.

Notice that, like most machine learning (ML) solutions,
\RankHow does not guarantee its results to be causal, that is, the returned function may not represent the true explanation.
However, in contrast to ML learning techniques, \RankHow's constraints
enable the user to actively explore alternative models and
fix undesirable outcomes in a targeted manner (see \Cref{ex:nba-mvp}).

While this paper focuses on scoring functions for the top-$k$
tuples, our approach easily generalizes to any subset of interest. For instance,
for a university ranked at position 50 that is interested in climbing the
ranks, \RankHow can provide a scoring function fit to the tuples ranked
at positions 30 to 50, simply by adjusting some program constraints.
Similarly, in addition to the ranking approximation error discussed,
\RankHow supports Kendall's Tau \cite{kendall-tau} and other measures
that are based on inversions, including variations that assign a
greater penalty to errors higher in the ranking.

\textbf{Why not just use linear regression?}
While linear regression and its variants like lasso can learn a linear
scoring function, they suffer from 2 major problems.
First, linear regression does not support flexible weight constraints.
Hence it may output a function with weights the user considers unrealistic
or undesirable.
Second, linear regression minimizes
a loss function over some response variable. Here the response variable
is the tuple's position in the ranking. (Recall that we do not know the
ground-truth scores, and they may not even exist.)
This is not the same as \emph{tuple-position-based optimization}:

\begin{example}[Prediction accuracy vs ranking accuracy]
Consider relation $R$ with 4 tuples $r_1,\ldots, r_4$, such that $r_i$
occupies position $i$ in the given ranking. To apply ML prediction techniques,
we can assign to $[r_1, r_2, r_3, r_4]$ numerical labels
$[4, 3, 2, 1]$---the tuple at position $i$ gets label $|R|-i+1$.
(W.l.o.g. the top-ranked tuples are those with the greatest scores.)
Now consider 2 competing models, one predicting scores $[8, 6, 2, 0]$
and the other predicting $[3, 2, 4, 1]$. The former has greater
squared error (26 vs 6) and absolute error (8 vs 4), but achieves
the perfect ranking, while the latter causes a total rank-position
error of 4 (and also 4 inversions), because it places $r_3$ at the top.
\end{example}

The example highlights the core drawback of existing
ML prediction techniques: They solve the problem of
accurately predicting a \emph{specific score value}, e.g., to assign
4 to $r_1$ and 3 to $r_2$, However, the true goal is to correctly
reproduce the given \emph{ordering} of the tuples, i.e., ranking $r_1$
above $r_2$, which can be achieved equally well by any function assigning
a greater score to $r_1$. The difference may appear subtle, but it
causes sub-optimal results for linear regression:

\begin{example}[Predicting scores vs predicting ordering]\label{ex:wrong_opt_goal}
Consider $R=\{(1,10000), (2,1000), (5,1), (4,10), (3,100)\}$ with attributes
$A_1, A_2$ and given rank vector $[1, 2, 3, 4, 5]$. \RankHow synthesizes
scoring function $0.99 A_1 + 0.01 A_2$, whose resulting scores
$[100.99, 11.98, 4.96, 4.06, 3.97]$ perfectly reproduce the ranking.
However, when applying linear regression to the corresponding $(A_1, A_2, y)$
triples, with $y$-vector $[5, 4, 3, 2, 1]$ as the ground-truth scores,
one obtains
$y=2.63608661 - 0.06685824 A_1 + 0.00025402 A_2$ and
$y=2.39465 + 0.00027 A_2$ for default settings and when requiring
coefficients to be non-negative, respectively. In both cases,
the resulting ranking becomes $[1, 2, 5, 4, 3]$, introducing an
error of 4 positions due to the swapping of the 3-rd and 5-th tuples.
\end{example}

Boosting techniques like AdaRank~\cite{07-sigir-xu} can in
theory ``retrofit'' position-based optimization to ML techniques optimizing
for another objective. However, they represent best-effort heuristics and
showed mixed results in our experiments.

\textbf{Why not use ML prediction techniques designed for
ordinal response variables or learning-to-rank?}
Techniques designed to learn rankings, in particular
\emph{ordinal classification}\footnote{These techniques are often
called ordinal \emph{regression}, even though they are closer to
classification techniques. We use ``ordinal classification''
to avoid confusion with ``ordinal regression'' as defined by
Srinivasan~\cite{76-jacm-srinivasan}.}
and \emph{learning-to-rank}, unfortunately cannot be meaningfully
applied to our problem (see \Cref{sec:related}).
In short, they either rely on training data containing many records (there is only a single training record for our problem:
the ranked dataset) or can be simplified to \LinearRegression in our context.

The only approach that does not require multiple training records or
guessing ``ground-truth'' scores and can be meaningfully applied
to our problem is the \textbf{ordinal regression} technique of
Srinivasan~\cite{76-jacm-srinivasan}. Still, it does not
support position-based optimization, and hence, like linear regression,
can prefer a sub-optimal scoring function.
(For more details, see \Cref{sec:related}.)
It also \emph{does not allow ties} in the given ranking.

We propose \RankHow to address the shortcomings of the state of the art,
making the following main contributions:
\begin{itemize}[nosep]
\item We formally define \OPT, the problem of synthesizing a given ranking
with a linear scoring function that satisfies constraints on attribute weights
(\Cref{sec:problem}). It also addresses the subtle problem of identifying
ties when scores are floating-point values, which are inherently imprecise.

\item We propose an MILP program that forms the foundation of \RankHow
(\Cref{sec:approach,sec:extension}). Even though MILP in general is NP-hard, we show
that our specific instance can be solved in polynomial time (PTIME)
in the number of input tuples. Interestingly, the PTIME algorithm
corresponds to a simpler instance of the sophisticated
approaches used by modern MILP solvers. Hence such an MILP solver
can generally be expected to solve \OPT at least as efficiently
as the PTIME algorithm. Our experiments
confirm orders of magnitude speedup compared to a PTIME algorithm
that is an extension of recent work~\cite{19-sigmod-asudeh-fair-ranking,24-icde-wang}.

\item To further improve scalability, we introduce heuristic \SymGD
(\Cref{sec:sym_gd}).
It represents a novel advanced version of gradient
descent that identifies the local optimum in the neighborhood around a seed point.

\item Experiments with real and synthetic datasets demonstrate that
\RankHow is practical and finds better linear scoring functions
than its competitors (\Cref{sec:experiments}).
\end{itemize}

\section{Problem Definition}
\label{sec:problem}

Our goal is to find linear scoring functions that approximate a
given ranking $\pi$ over a relation $R$, in the sense that when sorting the
$R$-tuples by score, their order is very similar to $\pi$.
We support a \emph{very general notion of ranking}: Each $R$-tuple
is assigned a position, which is a positive integer or symbol $\bot$,
such that (i) the lowest integer position is 1, (ii) there are no ``excessive gaps''
between integer positions, i.e., if a tuple is ranked at position $i$,
there should be at least $i-1$ tuples with a lower position, and
(iii) any tuple assigned $\bot$ is ranked at most as
high as the tuple with the greatest integer position.
Intuitively, assigning $\bot$ means that the order of lower-ranked
tuples does not matter. For instance, for $R = \{r_1, r_2, r_3, r_4, r_5, r_6\}$,
let $\pi=[\pi(r_1), \pi(r_2), \pi(r_3), \pi(r_4), \pi(r_5), \pi(r_6)]$
be the corresponding vector of tuple ranks. Valid rankings like
$[1, 2, 3, 4, \bot, \bot]$ and $[1, 1, 3, 3, \bot, \bot]$
express different situations with or without ties between the top-4;
and that $r_5$ and $r_6$ are not ranked higher than any top-4 tuple.
Rankings like $[2, 3, 4, 5, \bot, \bot]$ (does not start with 1)
and $[1, 1, 4, 4, \bot, \bot]$ (excessive gap between 1 and 4) are invalid. Formally:

\begin{definition}[given ranking]\label{def:given_ranking}
A ranking $\pi: R \rightarrow [1,\ldots, k, \bot]$ over dataset
$R = \{r_1,\ldots, r_n\}$ is a function that assigns a positive
integer or symbol $\bot$ to each $r \in R$, such that:
\begin{itemize}[nosep]
    \item $R_\pi(k) \subseteq R,\; |R_\pi(k)| = k$
    \item $\forall r \in R_\pi(k):\; \pi(r) \in [1,\ldots, k]$
    \item $\exists r \in R_\pi(k):\; \pi(r) = 1$
    \item $\forall r \in R_\pi(k):\; |\{r' \in R_\pi(k) \,|\, \pi(r') < \pi(r)\}| \ge \pi(r)-1$
    \item $\forall r \in R - R_\pi(k):\; \pi(r) = \bot$.
\end{itemize}
\end{definition}

\begin{table}[tb]
\caption{Notation}
\label{tab:notation}
\centering
\small
\begin{tabularx}{\linewidth}{@{\hspace{0pt}} >{$}l<{$}  @{\hspace{2mm}}  X @{}}
\hline
\textrm{Symbol}	& Definition 	\\
\hline
R               & Input dataset \\
n = |R|         & Number of tuples in $R$ \\
A_1,\ldots, A_m & Attributes of $R$ used for ranking \\
\score_W        & Linear scoring function over $A_1$,\ldots, $A_m$ \\
W=(w_1,\ldots, w_m) & Weight vector defining $\score_W$ \\
\pred           & Predicate constraining the choices of $W$ \\
\pi: R \rightarrow [1,\ldots, k, \bot] & Given ranking\\
\pi(r)          & Rank of $r \in R$ in $\pi$ \\
R_\pi(k)        & Top-$k$ tuples for given  ranking $\pi$ \\
\rank_W         & Ranking of $R$ induced by function $\score_W$ \\
\rank_W(r)      & Rank of $r$ based on scoring function $\score_W$ \\
R_W(k)          & Top-$k$ tuples for scoring function $\score_W$ \\
\epsilon        & Tie tolerance of a ranking \\
\tau            & Precision tolerance \\
\tau^+          & A value minimally greater than $\tau$ \\
\delta_{sr}     & Relationship indicator between $s, r \in R$ \\
\epsilon_1, \epsilon_2        & Thresholds for numerical imprecision \\
\hline
\end{tabularx}
\end{table}

We say that a tuple is \emph{ranked above} or \emph{ranked higher than}
another tuple iff it has a lower position.
\Cref{tab:notation} summarizes important notation.
A \emph{score-based ranking} over relation $R$ induces an
ordering in descending order based on the weighted sum of
numerical attributes $A_1$,\ldots, $A_m$ of $R$, i.e.,
\begin{equation}
\score_W(A_1, A_2,\ldots, A_m) = \sum_{i=1}^{m} w_i A_i, \label{eq:score_fct}
\end{equation}
where $W = (w_1,\ldots, w_m)$ and w.l.o.g., $w_i \ge 0, i=1,\ldots, m$
and $\sum_{i=1}^m w_1 = 1$.
We may omit $W$ from the function name when it is clearly implied.

\begin{definition}[score-based ranking]\label{def:score_based_rank}
Given a scoring function $\score_W$ on $R$, the \emph{rank} of a tuple
$r \in R$ is
\begin{align*}
\rank_W(r) =  \sum_{s \in R} \delta_{sr} + 1,
\end{align*}
where indicator $\delta_{sr}=1$ if
$\score_W(s) - \score_W(r) > \epsilon$,
for some $\epsilon \ge 0$, and $\delta_{sr}=0$ otherwise.
\end{definition}

There are two important aspects to note about the definition. First,
in the presence of score ties, all tied tuples have the same rank,
determined by the number of tuples with higher scores.
For instance, tuples $r_1,\ldots, r_4$
with scores 9, 6, 6, and 5, respectively, have ranks 1, 2, 2, and 4, respectively.
Second, a variable $\epsilon$ which is larger than 0 represents a ``safety gap'' that is necessary
when working with floating-point numbers, which are inherently
imprecise, including the ubiquitous floating-point arithmetic based on
the IEEE 754 standard. This means that 2 theoretically identical
scores $a, b$ may not satisfy $a=b$, and vice versa.
(In general, it is not meaningful to check floating-point numbers for equality.)
By setting $\epsilon$ greater than the numerical imprecision of the
floating-point type used by the programming language or solver, we can
meaningfully support ties of scores represented as floating-point numbers.
For example, for tuples $[r_1, r_2, r_3, r_4]$ with scores $[2.2, 2.1, 2.0, 1.5]$
and $\epsilon = 0.3$, the corresponding ranking is $[1, 1, 1, 4]$.
In a \emph{precise} environment, $\epsilon$ can be set to 0, so two tuples
are tied iff they have exactly the same score.
A larger $\epsilon$ establishes a larger gap between tuples that are not tied,
making the ranking more robust against small data changes.

Linear scoring functions may not be able to exactly reproduce
the given ranking. Hence we are interested in synthesizing a linear function
that minimizes ranking error. We focus on total position-based
error, i.e., if a top-$k$ tuple deviates by $x$ positions in the approximate ranking,
compared to the given ranking, then it contributes $x$ to the error:

\begin{definition}[position-based error]\label{def:position_error}
Given ranking $\pi$ and its approximation $\hat{\pi}$ over relation $R$.
The position-based error of $\hat{\pi}$ is
$\sum_{r \in R_\pi(k)} |\hat{\pi}(r) - \pi(r)|$.
\end{definition}

This measures directly how well the score-based ranking preserves the
tuple positions of the given ranking. Our approach generalizes
to other error measures.
We are now ready to define the optimization problem:

\begin{definition}[optimization problem \OPT]\label{definition:topkproblem}
Given a ranking $\pi$ and a conjunction $\mathcal{P}$ of linear constraints 
of type $\sum_{i=1}^m \alpha_i w_i \le \alpha_0$,
where $\alpha_i \in \mathbb{R}, i=0,\ldots,m$,
find the weight vector $W=(w_1,\ldots,w_m)$ that satisfies the constraints
and minimizes position-based error
$\sum_{r \in R_\pi(k)} |\rank_W(r) - \pi(r)|$.
\end{definition}

\begin{figure}[tb]
  \centering
  \includegraphics[width=0.75\linewidth]{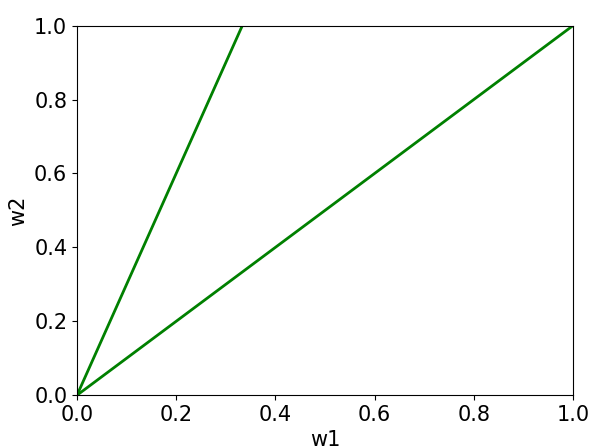}
  \caption{
  Weights resulting in ties (oblique lines) vs no ties (all others),
  and an example scoring function (star).
  }
  \label{fig:2dspace}
\end{figure}

\Cref{fig:2dspace} illustrates the problem in 2D space.
The diagonal lines correspond to scoring functions where some tuples'
scores are tied, separating regions without score ties.

\section{Exact Solution Using Linear Programs}
\label{sec:approach}

We propose an exact solution for \OPT based on an MILP program.
While MILP is NP-hard in general, we show that our instance can be solved
in PTIME in the number of input tuples. The PTIME algorithm
corresponds to a naive evaluation of the MILP program.

\subsection{MILP Solution for \OPT}

We propose an MILP program to exactly solve \OPT:
\begin{equation}
  \begin{aligned}\label{eq:best_LP_topk}
    \min & \quad  \sum_{r \in R_\pi(k)} \left| \pi(r) - 1 - \sum_{s \in R; s \neq r} \delta_{sr} \right|\\
    \text{s.t.} & \quad \mathcal{P}(w_1,\ldots,w_m)\\
    & \delta_{sr} = 1 \Rightarrow \sum_{i=1}^{m} w_i (s.A_i - r.A_i) \ge \epsilon_1,\\
    & \delta_{sr} = 0 \Rightarrow \sum_{i=1}^{m} w_i (s.A_i - r.A_i) \le \epsilon_2,\\
        & \hspace{25ex} r \in R_\pi(k), r \neq s \in R
  \end{aligned}
\end{equation}

Indicator $\delta_{sr}$ captures if tuple $s$ has a higher score than tuple $r$
($\delta_{sr}=1$) or not ($\delta_{sr}=0$). We use these indicators to compute
the score-based rank of each top-$k$ tuple $r$ as
$\hat{\pi}(r) = \sum_{s \in R; s \neq r} \delta_{sr}+1$ (\Cref{def:score_based_rank})
and thus the \textbf{ranking approximation error in the objective}
(\Cref{def:position_error}).

\begin{example}\label{example:indicator}
Consider $R(A_1, A_2, A_3)$ with tuples $r = (3, 2, 8)$, $s = (4, 1, 15)$,
and $t = (1, 1, 14)$, and given ranking $\pi[r, s, t] = [1, 2, \bot]$.
For $r$ the indicators are:
\begin{align*}
    &\delta_{sr} = \left( w_1 - w_2 + 7w_3 > 0\right), \\ &\delta_{tr} = \left(-2w_1 - w_2 + 6w_3 > 0\right),
\end{align*}
and since $\pi(r) = 1$, its contribution to the ranking error is
\begin{align*}
\left| \delta_{sr} + \delta_{tr} + 1 - \pi(r)\right| = \left| \delta_{sr} + \delta_{tr}\right|.
\end{align*}
Intuitively, since $r$ is at the top in the given ranking, any tuple beating it
in the score-based ranking increases its position error by 1.
For $s$ we can analogously derive error term
$\left| \delta_{rs} + \delta_{ts} - 1\right|$. Hence the overall error equals
$\left| \delta_{sr} + \delta_{tr}\right| + \left| \delta_{rs} + \delta_{ts} - 1\right|$.
\end{example}

Term $\mathcal{P}(w_1,\ldots,w_m)$ is a
\textbf{predicate that defines the conjunction of
all weight constraints}. In addition to the general constraints
$w_i \ge 0, i=1,\ldots, m$ and $\sum_{i=1}^m w_i = 1$, $\mathcal{P}$ can also
support flexible constraints\footnote{Other constraints, including the position
ranges discussed in \Cref{ex:nba-mvp}, can also be supported through linear
constraints involving the indicator variables, similar to the objective function.}
expressible as
$\sum_{i=1}^m \alpha_i w_i \le \alpha_0$, where $\alpha_i \in \mathbb{R}, i=0,\ldots,m$.
This way the solver is forced to only explore ``reasonable'' scoring functions,
or diverse scoring functions. 

The \textbf{indicator constraints} require a deeper analysis. We discuss
how to set $\epsilon_1$ and $\epsilon_2$ for a given $\epsilon$
(\Cref{def:score_based_rank}) in \Cref{sec:numerical_issues}.
Intuitively, these pre-set constants address floating-point
imprecision by (i) defining a range $\epsilon_2$ for
declaring scores ``tied'' and (ii) establishing a sufficiently large
gap $\epsilon_1$ for determining that one score definitely beats another.
The difference between $\epsilon_1$ and $\epsilon_2$ ensures that despite
numerical imprecision, the indicator can never have values 0 and 1 at the same time.
To this end, note that we want to establish the following:
\begin{equation}
  \begin{aligned}\label{eq:best_LP_topk_constraints_num}
    \sum_{i=1}^{m} w_i (s.A_i - r.A_i) \ge \epsilon_1 & \;\Rightarrow\; \delta_{sr} = 1\\
    \sum_{i=1}^{m} w_i (s.A_i - r.A_i) \le \epsilon_2 & \;\Rightarrow\; \delta_{sr} = 0\\
    \epsilon_2 < \sum_{i=1}^{m} w_i (s.A_i - r.A_i) < \epsilon_1 & \;\Rightarrow\; \mathtt{False},
  \end{aligned}
\end{equation}
where the last constraint establishes the ``safety gap'' to prevent
$\delta_{sr}$ from taking a wrong value because of numerical imprecision.

\begin{lemma}\label{lem:indicator_constr_num}
\Cref{eq:best_LP_topk_constraints_num} is logically equivalent to the indicator
constraints in \Cref{eq:best_LP_topk}
\begin{align*}
\delta_{sr} = 1 & \Rightarrow \sum_{i=1}^{m} w_i (s.A_i - r.A_i) \ge \epsilon_1\\
\delta_{sr} = 0 & \Rightarrow \sum_{i=1}^{m} w_i (s.A_i - r.A_i) \le \epsilon_2
\end{align*}
\end{lemma}
\begin{proof}
Consider the contrapositive of each of the constraints
in \Cref{eq:best_LP_topk_constraints_num}:
\begin{align*}
\delta_{sr} = 0 & \Rightarrow \sum_{i=1}^{m} w_i (s.A_i - r.A_i) < \epsilon_1\\
\delta_{sr} = 1 & \Rightarrow \sum_{i=1}^{m} w_i (s.A_i - r.A_i) > \epsilon_2\\
\mathtt{True} & \Rightarrow \left( \sum_{i=1}^{m} w_i (s.A_i - r.A_i) \le \epsilon_2\right.\\
& \left.\qquad \lor \sum_{i=1}^{m} w_i (s.A_i - r.A_i) \ge \epsilon_1 \right).
\end{align*}
From the second and third statements,
we derive $\delta_{sr} = 1 \Rightarrow \sum_{i=1}^{m} w_i (s.A_i - r.A_i) \ge \epsilon_1$,
and from the first and third statements,
$\delta_{sr} = 0 \Rightarrow \sum_{i=1}^{m} w_i (s.A_i - r.A_i) \le \epsilon_2$.
\end{proof}

In a \emph{precise} environment (e.g., an exact solver), we could
conceptually replace ``$\ge \epsilon_1$'' and ``$\le \epsilon_2$''
in \Cref{eq:best_LP_topk} with ``$> 0$'' and ``$\le 0$'',
corresponding to $\varepsilon = 0$ in \Cref{def:score_based_rank}.

\introparagraph{Program properties}
\Cref{eq:best_LP_topk} is a linear program with $k \cdot n$ binary
integer indicator variables and $m$ continuous weight variables plus weight constraints (at least 1).
We can directly solve it using an MILP solver
like Gurobi~\cite{gurobi}. SMT theorem provers like Z3~\cite{z3}
can be used if we convert the optimization problem to a series
of satisfiability problems, performing binary search to find
the smallest error value for which a satisfying assignment can be found.

\begin{figure}[tb]
  \centering
  \includegraphics[width=0.91\linewidth]{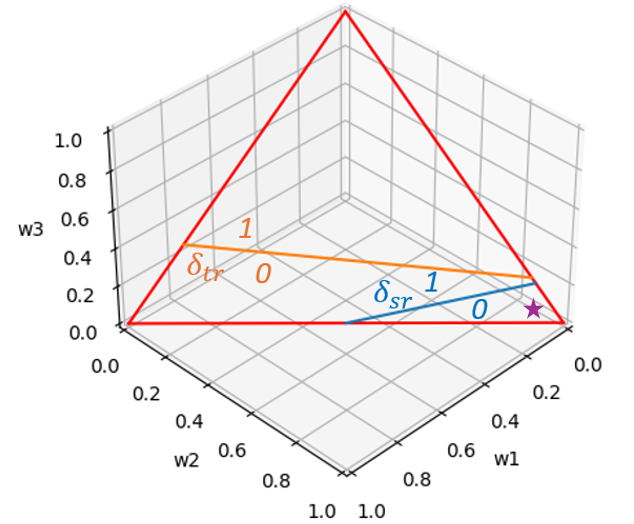}
  \caption{\Cref{example:space}: Solution space (2D triangle in 3D space)
  and indicator boundaries (2 lines in the triangle). The colored numbers show the
  indicator's value when selecting the weight from the corresponding side of
  the line.}
  \label{fig:3dspace}
\end{figure}

\begin{example}\label{example:space}
In \Cref{example:indicator}, the answer for \OPT is 0, because linear scoring
functions exist that perfectly rank $r$ before $s$ before $t$.
\Cref{fig:3dspace} illustrates the solution space, with the weights
$w_1, w_2, w_3$ assigned to the axes.
The red 2D triangle in 3D space represents the set of all $W$ where $w_1+w_2+w_3=1$.
Note that the star and the other 2 lines fall into the triangle.
Those lines show the boundaries for indicators $\delta_{tr}$ and
$\delta_{sr}$ (and hence also $\delta_{rs})$. Indicator $\delta_{ts}$
is not visible because it only intersects with the triangle at corner
point$(0, 1, 0)$: $s$ dominates $t$ and hence $\delta_{ts}$ must be 0.
The star represents a scoring function that perfectly recovers the
given ranking $\pi[r, s, t] = [1, 2, \bot]$, falling into the intersection
of half-spaces satisfying $\delta_{tr} = 0, \delta_{sr} = 0$. This intersection can intuitively be described
as the set of weight vectors with small $w_1$, large $w_2$, and
very small $w_3$.
\end{example}

\subsection{PTIME Solution and Why Not to Use It}
\label{sec:PTIME_alg}

\begin{theorem}[PTIME solution]\label{thm:PTIME}
The MILP program in \Cref{eq:best_LP_topk} can be solved in time complexity
that is polynomial in the number of input tuples in $R$.
\end{theorem}

\begin{proof}
The main proof idea is to replace the indicator variables in \Cref{eq:best_LP_topk}
one-by-one, until the MILP program is converted to a polynomial number
of LP programs, i.e., linear programs without integer variables.
Since LP can be solved in time that is polynomial in the number of variables
and constraints, solving a polynomial number of LP problems yields
polynomial time complexity overall.

For the LP reduction, consider a case where we have assigned a fixed
value---either 0 or 1---to each indicator
$\delta_{sr}, r \in R_\pi(k), r \neq s \in R$.
Then \Cref{eq:best_LP_topk} reduces to the LP problem of
returning any $(w_1,\ldots, w_m)$ that satisfies
\begin{equation}
  \begin{aligned}\label{eq:MILP_to_LP}
    & \mathcal{P}(w_1,\ldots,w_m)\\
    & \sum_{i=1}^{m} w_i (s.A_i - r.A_i) \ge \epsilon_1,\\
    & \hspace{15ex} \delta_{sr} = 1, r \in R_\pi(k), r \neq s \in R\\
    & \sum_{i=1}^{m} w_i (s.A_i - r.A_i) \le \epsilon_2,\\
    & \hspace{15ex} \delta_{sr} = 0, r \in R_\pi(k), r \neq s \in R
  \end{aligned}
\end{equation}
Intuitively, when fixing the value of $\delta_{sr}$, its
2 indicator constraints in \Cref{eq:best_LP_topk} are replaced by the
corresponding inequality. Since there are no integer variables in
\Cref{eq:MILP_to_LP}, we are left with an LP problem with
$m$ variables and $kn$ constraints, which can be solved in time
polynomial in $m, kn$.

Generating all $2^{kn}$ combinations of 0/1 assignments to the $kn$
indicators produces an exponential number of LP problems. We avoid this
by exploiting a well-known result from computational geometry
that the number of distinct partitions
generated by $x$ hyperplanes in $d$-dimensional space is $\O(x^d)$ \cite{geometry}.
Note that for each indicator $\delta_{sr}$, the corresponding hyperplane
is defined by the score-difference formula $\sum_{i=1}^{m} w_i (s.A_i - r.A_i) = 0$.
Since our problem has $kn$ indicators and $m$ dimensions,
the number of possible
partitions is limited to $\O(k^m n^m)$, which is polynomial in $n$, the number of
tuples in $R$.\footnote{Since the $m$ weights add up to 1, there are only $m-1$ independent
variables and thus only $m-1$ dimensions. For simplicity and readability,
we use $m$ in the formulas, noting that the implementation depends on $m-1$.}

In order to generate all distinct partitions defined by the $kn$ hyperplanes
in time $\O(k^m n^m)$, we can use standard depth-first search (DFS) with backtracking or breadth-first search (BFS).
To see how the algorithm works, consider a node $N$ in the corresponding tree
of states and let $\mathcal{I}$ denote the conjunction of all
inequalities of type $\sum_{i=1}^{m} w_i (x.A_i - y.A_i) \ge \epsilon_1$
or $\sum_{i=1}^{m} w_i (x.A_i - y.A_i) \le \epsilon_2$, depending on the value of
each indicator $\delta_{xy}$ the algorithm has branched on along the path from the root
to node $N$. To expand from $N$, the algorithm picks one of the remaining indicators,
say $\delta_{rs}$. It checks the relationship between $N$'s feasible region
(defined by $\mathcal{I}$) and the hyperplanes
$\mathcal{H}_1 = \sum_{i=1}^{m} w_i (s.A_i - r.A_i) \ge \epsilon_1$
and $\mathcal{H}_0 = \sum_{i=1}^{m} w_i (s.A_i - r.A_i) \le \epsilon_2$ for
$\delta_{rs}$. Formally, the algorithm first checks
if $\mathcal{I} \land \mathcal{H}_0$ is satisfiable.
If so, it adds a child node $C_0$ to $N$, whose feasible region
constraint is $\mathcal{I} \land \mathcal{H}_0$. Then it analogously
checks $\mathcal{I} \land \mathcal{H}_1$ and, if satisfiable, adds
another child node $C_1$ with feasible region constraint
$\mathcal{I} \land \mathcal{H}_1$. Both satisfiability checks are
LP programs (no integer variables) and hence take polynomial time.
Notice that since $\mathcal{H}_0 \land \mathcal{H}_1$ is unsatisfiable
because $\epsilon_1 > \epsilon_2$, the algorithm cannot encounter
the same partition more than once. And due to the $\O(k^m n^m)$ upper
bound on the number of possible partitions, the number of nodes reached
by the search algorithm is $\O(k^m n^m)$ as well.
(Intuitively, many nodes only have
a single child, because their feasible region lies completely
above or completely below the indicator's hyperplane.)
\end{proof}

Since the proof of \Cref{thm:PTIME} is constructive, we could use the corresponding algorithm to solve \OPT in time polynomial in the number
of tuples in $R$.
\textbf{While a guaranteed PTIME solution may appear preferable compared to
the theoretically NP-hard world of MILP, we argue that holistically
solving \Cref{eq:best_LP_topk} with an MILP solver is the better approach.}
This hypothesis is confirmed by our empirical evaluation in \Cref{sec:case_study}. Modern
MILP solvers already rely on advanced backtracking algorithms that allow them to use 
results from one part of the search space to rule out the existence of solutions
in other parts of the search space. And these techniques often lead to 
significant improvements compared to simple backtracking~\cite{davey2002efficient},
where sub-problems for the different partitions defined by the
hyperplanes are fed piece-meal to separate LP instances.
In short, the PTIME approach in the proof of \Cref{thm:PTIME},
as well as similar techniques for related ranking
problems~\cite{19-sigmod-asudeh-fair-ranking, 24-icde-wang},
correspond to simple algorithms that lack
the clever cross-branch information passing and other
sophisticated strategies employed by modern MILP solvers.

\section{Symbolic Gradient Descent}
\label{sec:sym_gd}

We introduce \SymGD and discuss its design decisions.

\subsection{\SymGD Overview and Algorithm}

\begin{algorithm}[tb]
    \caption{Symbolic Gradient Descent}
    \label{alg:sgd}
    \begin{algorithmic}[1]
        \REQUIRE Seed point $W_0$, cell size $c$, $i=0$
        \ENSURE Winning weight vector $W_i$
        \REPEAT
            \STATE $i$ = $i + 1$
            \STATE $W_i =$ solve($W_{i-1}, c$)
        \UNTIL{error($W_i$) = error($W_{i-1}$)}
        \RETURN $W_i$
    \end{algorithmic}
\end{algorithm}

\SymGD, like the well-known gradient-descent approaches,
starts from a seed point, here an initial weight vector $W_0$,
and then moves toward a local minimum (of ranking approximation error).
However, it is more like ``gradient descent on steroids'':
Instead of moving toward a local optimum, \SymGD actually finds the true
local optimum within a cell of size $c$ around the seed point.
This means that it does not require a differentiable error function,
but it needs an appropriate solver, such as an MILP solver in the case of \OPT.
Upon finding the local optimum, the cell shifts accordingly and the process
continues in the new neighborhood. \Cref{alg:sgd} summarizes the basic idea.
Function solve($W, c$) runs the MILP solver for \Cref{eq:best_LP_topk},
with the additional constraints that $W$ falls into the cell of size $c$
around seed point $W$: $\max(w_i - c / 2, 0) \le w_i \le min(w_i + c / 2, 1)$
for $i \in \{1,\ldots, m\}, 0 < c < 2$.

\SymGD exploits the structure of \Cref{eq:best_LP_topk}, which ensures
that \textbf{execution time drops dramatically as cell size decreases}.
Intuitively, the smaller a cell in weight space, the fewer of the
indicator hyperplanes $\sum_{i=1}^{m} w_i (s.A_i - r.A_i) = \epsilon_1$
and $\sum_{i=1}^{m} w_i (s.A_i - r.A_i) = \epsilon_2$ intersect it,
meaning the corresponding indicator value is uniquely determined
and hence the indicator becomes a constant.
In the extreme, for a cell not intersected by any of the hyperplanes,
\Cref{eq:best_LP_topk} reduces to the LP problem in \Cref{eq:MILP_to_LP}.
Stated differently, in a sufficiently small cell, the complexity of
\OPT drops by a factor of $\O(k^m n^m)$, which can be \textbf{several orders of magnitude}
for large input.

\subsection{How to Select the Seed Point?}

\RankHow currently supports 2 main strategies for selecting seed points.
The first relies on a fast heuristic like standard linear regression
or ordinal regression, which both treat the given tuple ranks as labels.
Even though both optimize for the wrong loss measure, that measure
is correlated with rank-position error, thus in our experiments especially
ordinal regression often identified good weight vectors that
\SymGD was able to improve.
The second strategy first divides the space into $(1 / c) ^ m $ cells and choose the cell with the smallest lower bound of the error.
The approach to determining the lower bound of a cell is based on the following insight: Consider any
regularly-shaped region in the weight space, e.g., an $m$-dimensional
hyperrectangle $C$. For each indicator hyperplane
$\sum_{i=1}^{m} w_i (s.A_i - r.A_i) = \epsilon_1$
and $\sum_{i=1}^{m} w_i (s.A_i - r.A_i) = \epsilon_2$,
we can determine if $C$ is intersected by the hyperplane, fully above it, or
fully below it. For the latter two, the corresponding indicator value
is fixed. Therefore, for each $r \in R$, we can derive a lower bound (also an upper bound) for the number of tuples that beat $r$. 
This in turn
determines a lower and an upper bound for the ranking-approximation error
for any weight vector in $C$.

\subsection{How to set the cell size?}

\begin{algorithm}[tb]
    \caption{Adaptive Symbolic Gradient Descent}
    \label{alg:sgd_cell}
    \begin{algorithmic}[1]
        \REQUIRE Seed point $W_0$, starting cell size $c$, $i=0$
        \ENSURE Winning weight vector $W_i$
        \REPEAT
            \REPEAT
            \STATE $i$ = $i + 1$
            \STATE $W_i =$ solve($W_{i-1}, c$)
            \UNTIL{error($W_i$) = error($W_{i-1}$) or timeout $t_{\text{total}}$}
        \STATE $c$ = 2$c$
        \UNTIL{timeout $t_{\text{total}}$}
        \RETURN $W_i$
    \end{algorithmic}
\end{algorithm}

Larger cells prevent getting stuck in a poor local optimum, but they increase
running time for the solver. Since all weights are between 0 and 1, cell size
can be any value from 0 to 2. Picking too large a cell size can
eat up the entire time budget $t_{\text{total}}$ for a single (unsuccessful)
iteration of \Cref{alg:sgd}. We therefore propose the adaptive approach
shown in \Cref{alg:sgd_cell}: Start with a small cell size and then
increase it whenever the algorithm gets stuck in a local optimum.
(Note that for larger $m$, we can modify \Cref{alg:sgd_cell} to increase
cell size in only some of the dimensions.)

\section{Details and Extensions}
\label{sec:extension}

\subsection{Dealing with Numerical Imprecision}
\label{sec:numerical_issues}

We discuss how to set constants $\epsilon_1$ and $\epsilon_2$
in \Cref{eq:best_LP_topk}. They are needed to support ties and avoid
incorrect results caused by numerical imprecision when working with
floating-point numbers.
Numerical issues caused by the limited precision of real-world data types
are well-known, but rarely receive attention in database research.
This is not surprising, because when one cares about a value in isolation,
e.g., an aggregate over a relation or the probability returned by a prediction model,
a small error does not change the nature of the output.
However, \textbf{in the context of ranking, even a tiny imprecision can
alter the tuple ordering}.
Consider tuples $r_1$ and $r_2$ whose scores 1.1 and 1.0 could each deviate
from their true value by up to 0.1. Even though $r_1$ nominally looks like
the winner, when taking the imprecision into account, the \emph{true}
score of $r_2$ may be greater.
Furthermore, Gurobi, like other high-performance MILP solvers,
considers a constraint over floating-point variables satisfied,
whenever it is ``close enough.'' For instance, Gurobi would
consider the theoretically unsatisfiable constraint
$x \le 0.0 \land x \ge 10^{-10}$ as satisfied for $x=0.0$~\cite{gurobi-tolerances}.
While precise alternatives often exist, e.g., exact MILP solvers like
exact SCIP~\cite{scip}, they are rarely used in practice due to poor
performance compared to the ``imprecise'' counterparts.

Let $\tau$ be the smallest value, called the precision tolerance, such that
the following holds for all indicator constraints in \Cref{eq:best_LP_topk}:
If the solver considers $\score_W(s) - \score_W(r) \ge \varepsilon_1$
satisfied, then $\score_W(s) - \score_W(r) \ge \varepsilon_1 - \tau$.
Analogously, if the solver considers $\score_W(s) - \score_W(r) \le \varepsilon_2$
satisfied, then $\score_W(s) - \score_W(r) \le \varepsilon_2 + \tau$.
The following lemmas specify how to set $\epsilon_1, \epsilon_2$
in \Cref{eq:best_LP_topk} depending on $\tau$ so that numerical
imprecision does not result in the solver admitting an incorrect solution.

\begin{lemma}\label{lem:eps-diff}
Setting $\epsilon_1 > \epsilon_2 + 2 \tau$
ensures that $\score_W(s) - \score_W(r) \ge \epsilon_1$
and $\score_W(s) - \score_W(r) \le \epsilon_2$
cannot both be considered true by the solver
for any pair $r, s$.
\end{lemma}
\begin{proof}
By definition of $\tau$, if the solver considers both
$\score_W(s) - \score_W(r) \ge \epsilon_1$ and
$\score_W(s) - \score_W(r) \le \epsilon_2$ be satisfied,
then $\score_W(s) - \score_W(r) \ge \varepsilon_1 - \tau$ and
$\score_W(s) - \score_W(r) \le \varepsilon_2 + \tau$.
For the latter to be false for all $r, s$, it is sufficient to ensure
$\varepsilon_2 + \tau < \varepsilon_1 - \tau$.
\end{proof}

\begin{lemma}\label{lem:eps-2-ties}
For a given $\epsilon$, setting $\epsilon_2 \ge \epsilon - \tau$
is necessary to guarantee that $|\score_W(s) - \score_W(r)| \le \epsilon$
implies $\delta_{sr} = \delta_{rs} = 0$.
\end{lemma}
\begin{proof}
For the sake of contradiction, assume $\epsilon_2 < \epsilon - \tau$. Now consider
$\epsilon_2 + \tau < \score_W(s) - \score_W(r) \le \epsilon$. For score differences
in this range, $\score_W(s) - \score_W(r) \le \epsilon_2 + \tau$ is false
and hence the solver does not consider $\score_W(s) - \score_W(r) \le \epsilon_2$
satisfied. Thus $\delta_{sr} = 0$ is not true, even when accounting for
the numerical tolerance.
\end{proof}

Using $\varepsilon$ from \Cref{def:score_based_rank}, we propose
$\epsilon_2 = \epsilon - \tau$ and $\epsilon_1 = \epsilon + \tau^+$,
where $\tau^+$ is a value minimally greater than $\tau$.
Based on \Cref{lem:eps-diff}, since $\epsilon_1 - \epsilon_2 = \tau + \tau^+ > 2\tau$,
\emph{numerical issues are avoided}, i.e., $\delta_{sr}$ cannot be 0 and 1 at the same time.
Setting $\epsilon_1 = \epsilon + \tau^+$ ensures that 
$\score_W(s) - \score_W(r) > \epsilon$ if $\delta_{sr} = 1$.
Based on \Cref{lem:eps-2-ties}, setting $\epsilon_2 = \epsilon - \tau$ ensures that
$\delta_{sr} = 0$ only if $\score_W(s) - \score_W(r) \le \epsilon$.
In this way, the \emph{imprecise solver does not produce false positives},
i.e., ``solutions'' that fail verification. Here verification means that
the values of the indicator variables are consistent with a score-based
ranking determined by a program using precise arithmetic, e.g., BigDecimal in Java.

We propose a heuristic to find $\tau$ using binary search: Given $\hat{\tau}$ an estimated value of $\tau$, we run the solver
on \Cref{eq:best_LP_topk}. If there are numerical problems, then we continue with a
larger value; otherwise with a smaller value.
Here we detect numerical problems by discovering a false positive through verification
of the solution using a program with precise arithmetic as discussed above.
Note that setting $\hat{\tau}$ too large, i.e., $\hat{\tau} > \tau$, eliminates
the range
$\epsilon - \hat{\tau} + \tau \le \score_W(s) - \score_W(r) \le \epsilon + \hat{\tau} - \tau$
from the solution space, which can lead to false negatives:
The solver cannot find a scoring function for which the score difference of 2 tuples
falls into this range.

\subsection{Removing Indicators of Dominators and Dominatees}

Removing indicator variables for all dominator-dominatee pairs from
\Cref{eq:best_LP_topk} reduces solver time and space cost, without
impacting result quality. 
A tuple $s$ dominates $r$ iff it has greater values for all ranking attributes
$A_1, \ldots, A_m$ compared to $r$. For such a pair,
we set $\delta_{sr} = 1$ and $\delta_{rs} = 0$, replace the indicator
variables with the corresponding value in the objective function, and remove
the corresponding indicator constraints from \Cref{eq:best_LP_topk}.
Finding all dominator-dominatee pairs for top-k tuples takes time $\O(kn)$, which is lower than
the time complexity of solving the linear program. Hence we always apply
this pre-processing step.

\section{Experiments}
\label{sec:experiments}
We show that our techniques are practical and
compare them against related work in terms of both effectiveness and efficiency.
For \SymGD, we demonstrate good approximation quality and quantify the tradeoff
between running time and result quality.
Using 39 different combinations of datasets and given rankings, we evaluate
the effects of parameters ($n$, $m$, $k$), the data distribution and
the ranking function used to generate the given ranking, demonstrating
the applicability of \RankHow to realistic non-linear ranking functions.
The source code and data are available at \url{https://github.com/northeastern-datalab/rankhow}.

\subsection{Experimental Setup}
\label{sec:setup}

\introparagraph{Environment and implementation}
All experiments are executed on an Ubuntu 20 Linux server with an
Intel Xeon E5-2643 CPU and 128GB RAM. 
We implemented our solvers and optimizers in Java 11, using the Java
libraries of the leading commercial optimizer Gurobi~\cite{gurobi}, version 9.5.2.
For Gurobi, we used the default configuration for multi-threaded execution,
which on our machine utilized 8 cores with up to 16 threads.
\SymGD by default starts from a seed point generated by \OrdinalRegression.
We use \Cref{alg:sgd_cell} starting with an initial cell size $10^{-4}$ given a timeout threshold and \Cref{alg:sgd} otherwise.
For numerical stability, we set $\epsilon = 5 \times 10 ^{-5}, \epsilon_1 = 10^{-4}, \epsilon_2 = 0$
for the NBA dataset, $\epsilon = 5 \times 10^{-3}, \epsilon_1 = 10^{-2}, \epsilon_2 = 0$
for the CSRankings dataset, and $\epsilon = 5 \times 10^{-6}, \epsilon_1 = 10^{-5}, \epsilon_2 = 0$ for all synthetic datasets.
All solutions are verified.
Competitors \LinearRegression and \Adarank are implemented in Python 3.8.10.

\introparagraph{Competitors}
We implemented all known competitors:
\Tree, \OrdinalRegression, \LinearRegression, \Adarank and \Sampling.
\Tree is the arrangement-tree-based PTIME 
algorithm\footnote{The algorithm uses BFS for tree construction.}~\cite{19-sigmod-asudeh-fair-ranking}.
\OrdinalRegression refers to ordinal regression, which optimizes for a
minimum score penalty \cite{76-jacm-srinivasan}. We extended it to support
ties and numerical imprecision, which can be turned off to obtain the
original technique.
We also apply the AdaRank algorithm of Xu and Li \cite{07-sigir-xu} to our problem,
which is a boosting technique originally designed for ranking documents
in response to search queries.
The algorithm trains multiple weak rankers---single features are used
as weak rankers in their paper---on the training data (tuples from input $R$)
and updates the weight distribution according to the performance on each tuple.
The measure of prediction quality is the ranking error of a tuple.
i.e., by how many positions its ranking is off when sorting according to the scoring function.
We refer to this implementation as \Adarank.
\LinearRegression and
\Adarank are the most meaningful adaptation to our problem
of existing explainable learning-to-rank
techniques that return linear scoring functions (see \Cref{sec:related}).

\introparagraph{Datasets and corresponding given rankings}
Two real datasets and nine large synthetic datasets of three different distributions (three for each) are used in the experiments.
The given rankings for the real datasets are generated by \emph{complex non-linear ranking functions} or aggregations of voting.
On each of the synthetic datasets, we use four different ranking functions (non-linear ones with exponents up to 5) to generate the given ranking.
Each combination of a dataset and a given ranking presents a different input
problem for \RankHow.
All techniques receive only the resulting ranking, i.e., the true
scores are hidden from them.

Like previous work on related problems \cite{23-pvldb-chen,10-icde-vlachou,12-icde-he},
we use the \NBA data~\cite{nba}, a real dataset containing 22840 tuples with statistics
of all NBA players from seasons 1979/80 to 2022/23.
Each tuple represents a \emph{player-season combination}---uniquely identified by
the PLR attribute, which consists of player name, age and team.
The default ranking attributes are the player's average statistics during
a season: PTS, REB, AST, STL, BLK, FG\%, 3P\%, and FT\%.
For players with identical statistics, we only keep one of them.
The given ranking is generated by either
the MVP ranking or by using MP (Minutes Played) * PER (Player Efficiency Rating),
whose computation involves a complicated formula and a few additional
attributes that are not included in the basic ranking attributes~\cite{nba-per}.

We also study the CSRankings data \cite{csrankings-github},
which contains 628 tuples with the publication numbers in 27 areas of computer science.
Compared to the \NBA data, the CSRankings dataset has fewer tuples but many more attributes.
The default ranking is obtained from the CSRankings website \cite{csrankings}.

To show that \SymGD scales to larger data, we generated nine synthetic
datasets of three different distributions (\textit{uniform}, \textit{correlated},
and \textit{anti-correlated}) with 1 million tuples each.
They allow us to explore the impact of correlations between the ranking
attributes. For the uniform data, values for each ranking attribute
are generated uniformly at random.
In the correlated dataset, a tuple with a high (low) value in one ranking
attribute is likely to also have high (low) values for the others.
In the anti-correlated dataset, a tuple with a high (low) value in one ranking
attribute is likely to have high (low) values for half of the other
attributes and low (high) values for the other half.
This pattern of generating synthetic data of different distributions
dates back to~\cite{synthetic}.

\introparagraph{Parameters}
\Cref{tab:parameter} shows the parameter values explored for \Cref{sec:exact,sec:approximation,sec:scalability}, with the default
setting in bold. We generally use more challenging settings for the
problems that are more efficiently solvable.

\begin{table*}[tb]
    \centering
    \caption{Experimental parameter settings}
    \begin{tabular}{|c|c|c|c|}
    \hline
    Parameter & NBA(\Cref{sec:exact}) & CSRankings(\Cref{sec:exact}) & Synthetic (\Cref{sec:scalability})\\
    \hline
    Length of the given top ranking ($k$) & 2, 3, 4, \textbf{5}, 6 & \textbf{5}, 10, 15, 20, 25 & 5, \textbf{10}, 15, 20, 25\\  
    \hline
    Number of tuples ($n$) & \makecell{5000, 10000, 15000, \\ 20000, \textbf{22840}} & \makecell{100, 200, 300, 400, \\ 500, 600, \textbf{628}} & 1000000\\ 
    \hline
    Number of attributes ($m$) & 4, \textbf{5}, 6, 7, 8 & \textbf{5}, 10, 15, 20, 25, 27 & 5\\
    \hline
    Distribution & Real-world & Real-world & Uniform, correlated, anti-correlated\\
    \hline
    Given ranking ($\pi$) & PER \cite{nba-per} & Default CSRankings \cite{csrankings} & $\sum{A_i ^{2}}$, $\mathbf{\sum{A_i ^{3}}}$, $\sum{A_i ^{4}}$, $\sum{A_i ^{5}}$ \\
    \hline
    \end{tabular}
    \label{tab:parameter}
\end{table*}

\subsection{Case Study: NBA MVP}
\label{sec:case_study}

We study the ranking for the 2022-23 NBA MVP award. The given ranking is based on the panel's selections (see \Cref{ex:nba-mvp}).
13 players received at least 1 vote, with the last two being tied;
there are 8 ranking attributes (\Cref{sec:setup}).
In 1.6 sec, \RankHow returns the optimal result.
The corresponding score-based ranking is
[1, 3, 4, 4, 2, 6, 7, 8, 9, 10, 11, 12, 13], with the numbers indicating the
player's position in the given ranking. Hence total position error is 6,
approximating the given ranking well. 
When applying \Tree to the same problem,
it took more than 16 hours (!) to return a weight vector with
an error of 9. After applying our $\epsilon_1$ construction to \Tree,
it became faster (because $\epsilon_1$ helped eliminate many tree nodes),
taking 36 minutes, and reducing the error to 7. It did not reach error 6, because
without adding our $\epsilon_2$ construction, \Tree is unlikely to sample
from a partition a weight vector that results in ties.
In summary, the original \Tree algorithm is more than 35,000 times slower
than \RankHow, and even after reducing tree size through $\epsilon_1$,
it still was 1000 times slower. 
This experimental result supports our argument in \Cref{sec:PTIME_alg} for
using the MILP solver instead of the PTIME algorithm.
Therefore, we do not include \Tree in the remaining experiments.

\subsection{Exact \OPT}
\label{sec:exact}

\begin{figure*}[tbp]
\captionsetup[subfigure]{justification=centering}
    \centering
    \begin{subfigure}[h]{0.29\textwidth}
        \centering
        \includegraphics[width=1.02\textwidth]{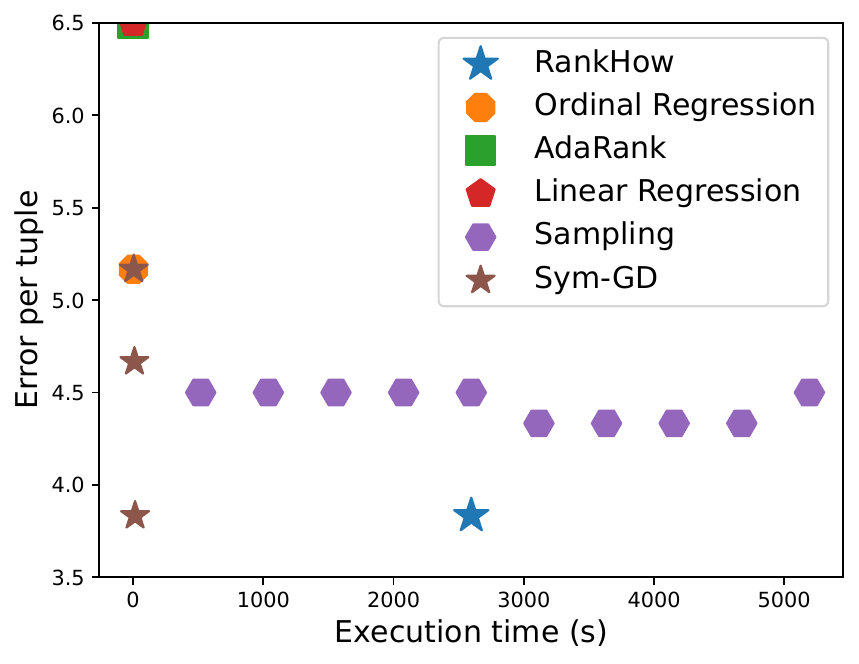}
        \caption{\NBA data, big picture}
        \label{fig:experiment_approaches}
    \end{subfigure}
    ~
    \begin{subfigure}[h]{0.29\textwidth}
		\centering
		\includegraphics[width=\textwidth]{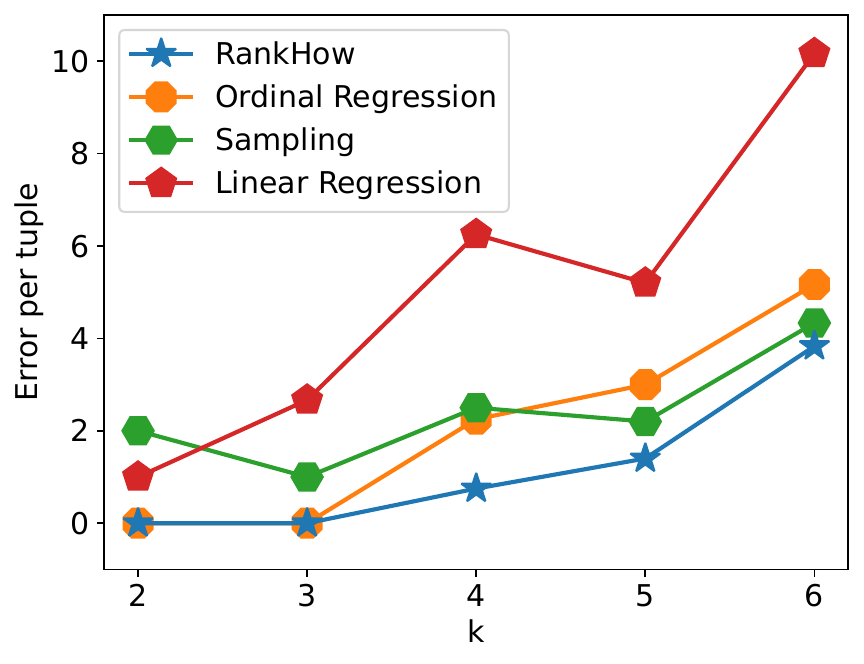}
        \caption{\NBA data}
        \label{fig:experiment_nba_k}
    \end{subfigure}
    ~
    \begin{subfigure}[h]{0.29\textwidth}
        \centering
        \includegraphics[width=\textwidth]{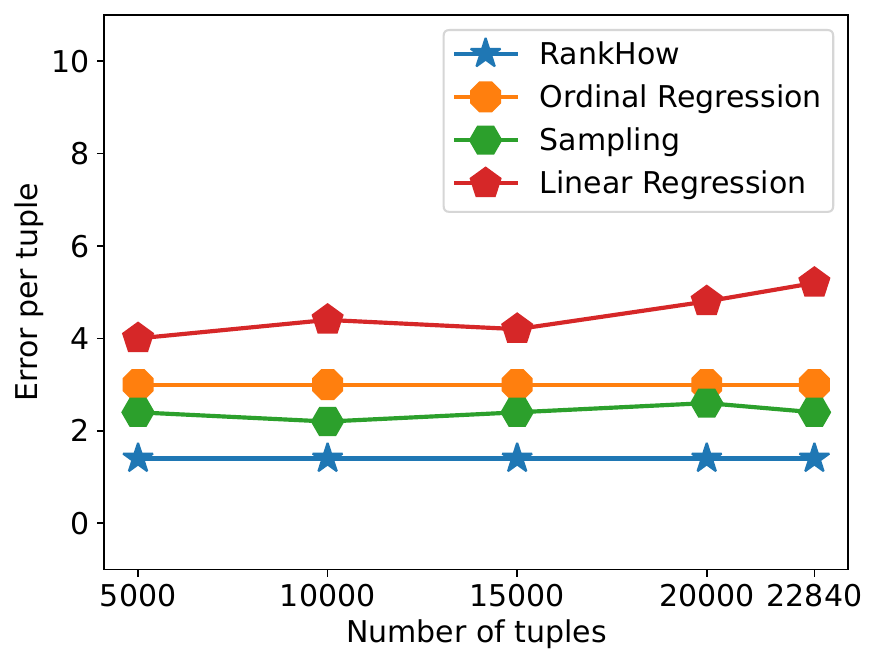}
        \caption{\NBA data}
        \label{fig:experiment_nba_n}
    \end{subfigure}

    \begin{subfigure}[h]{0.29\textwidth}
        \centering
        \includegraphics[width=\textwidth]{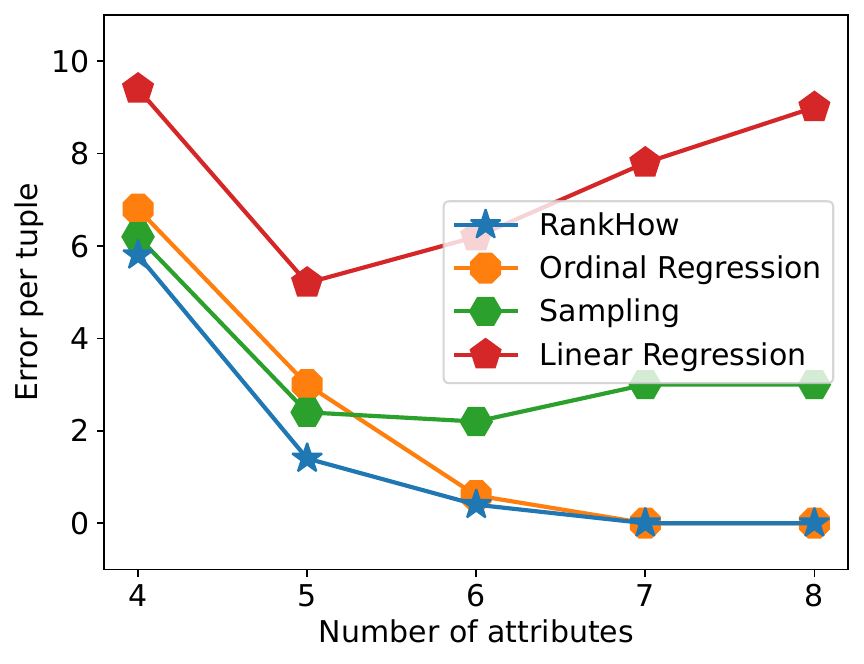}
        \caption{\NBA data}
        \label{fig:experiment_nba_m}
    \end{subfigure}
    ~
    \begin{subfigure}[h]{0.29\textwidth}
		\centering
		\includegraphics[width=\textwidth]{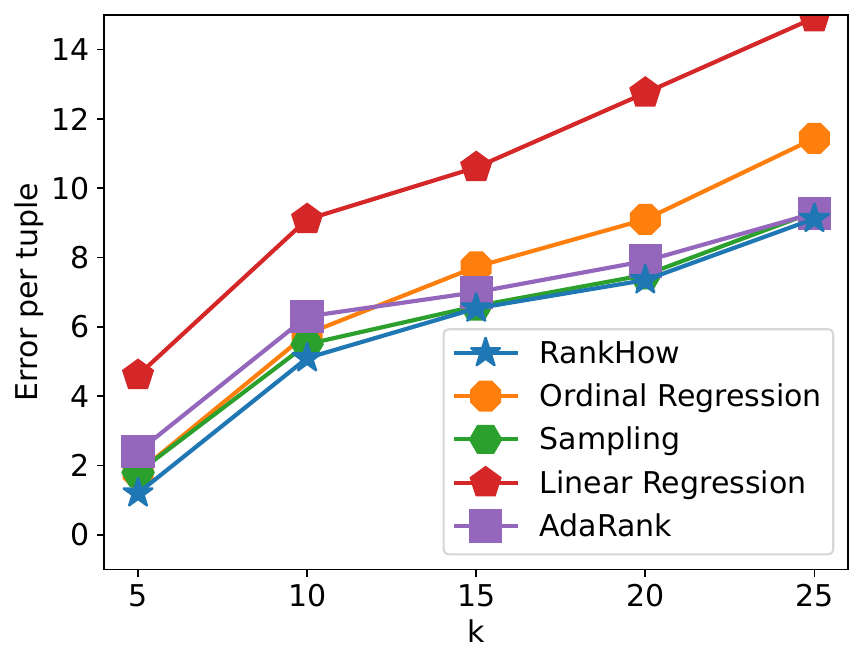}
        \caption{CSRankings data}
        \label{fig:experiment_csrankings_k}
    \end{subfigure}
    ~
    \begin{subfigure}[h]{0.29\textwidth}
        \centering
        \includegraphics[width=\textwidth]{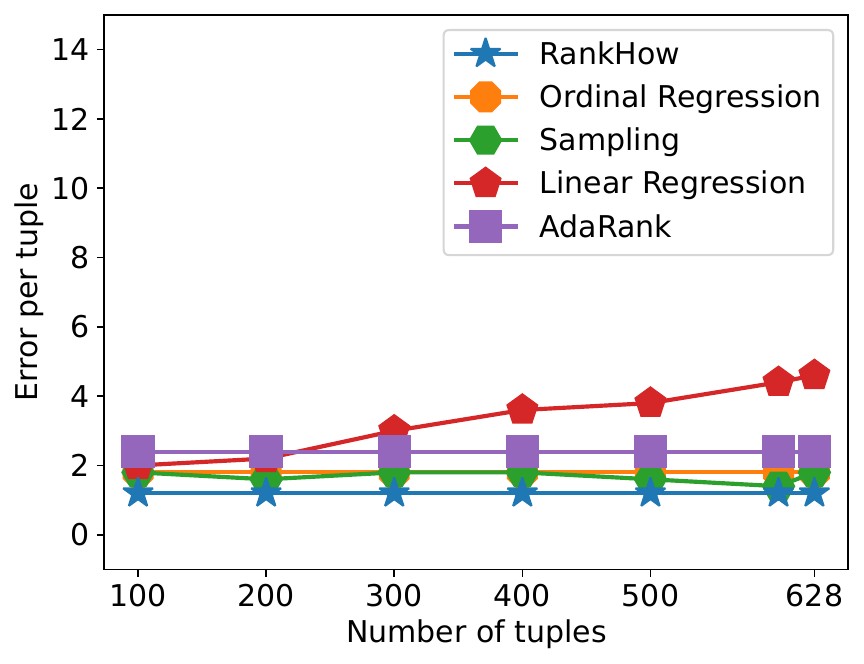}
        \caption{CSRankings data}
        \label{fig:experiment_csrankings_n}
    \end{subfigure}
    
    \begin{subfigure}[h]{0.29\textwidth}
        \centering
        \includegraphics[width=\textwidth]{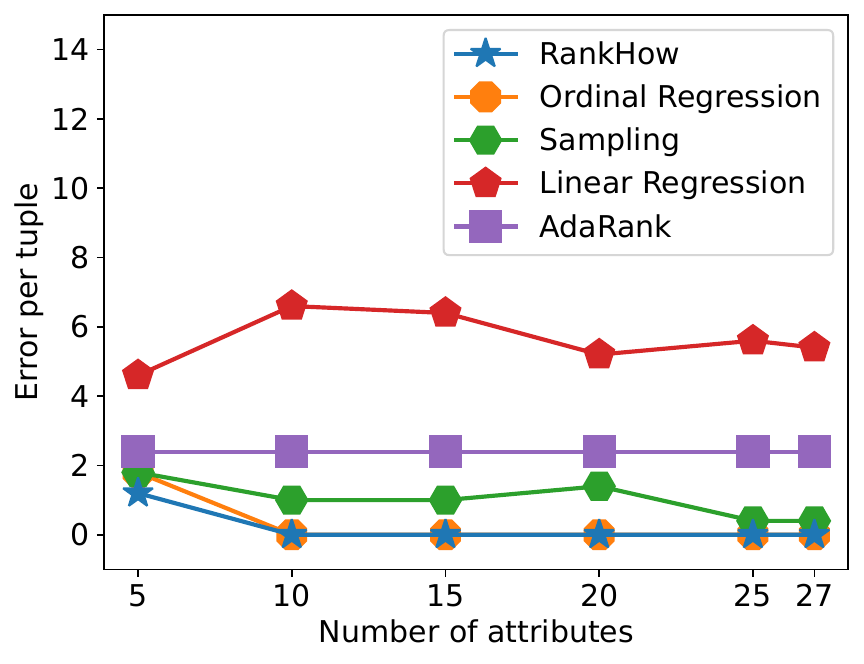}
        \caption{CSRankings data}
        \label{fig:experiment_csrankings_m}
    \end{subfigure}
    ~
    \begin{subfigure}[h]{0.29\textwidth}
        \centering
        \includegraphics[width=1.02\textwidth]{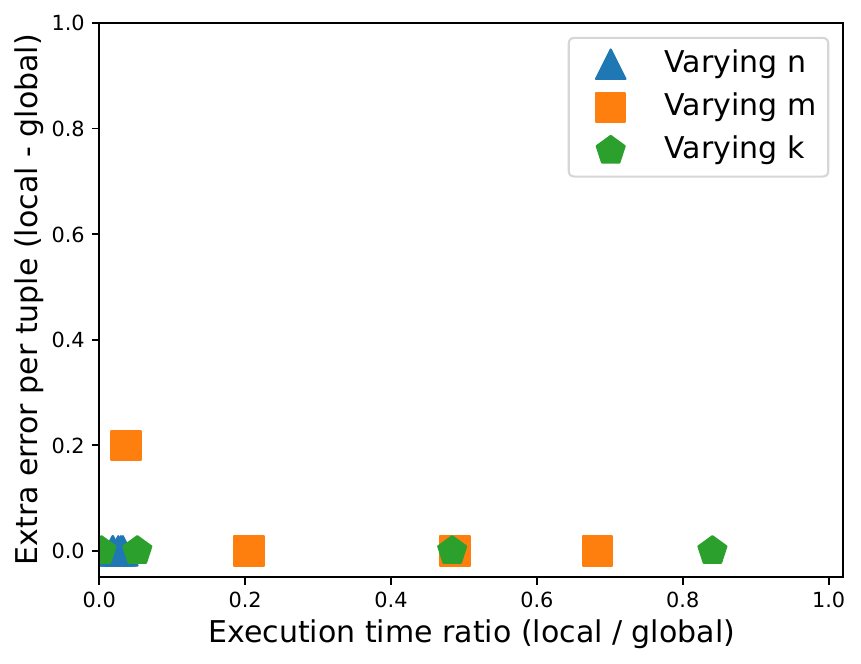}
        \caption{\NBA data, approximation quality}
        \label{fig:experiment_approximation}
    \end{subfigure}
    ~
    \begin{subfigure}[h]{0.29\textwidth}
        \centering
        \includegraphics[width=1.1\textwidth]{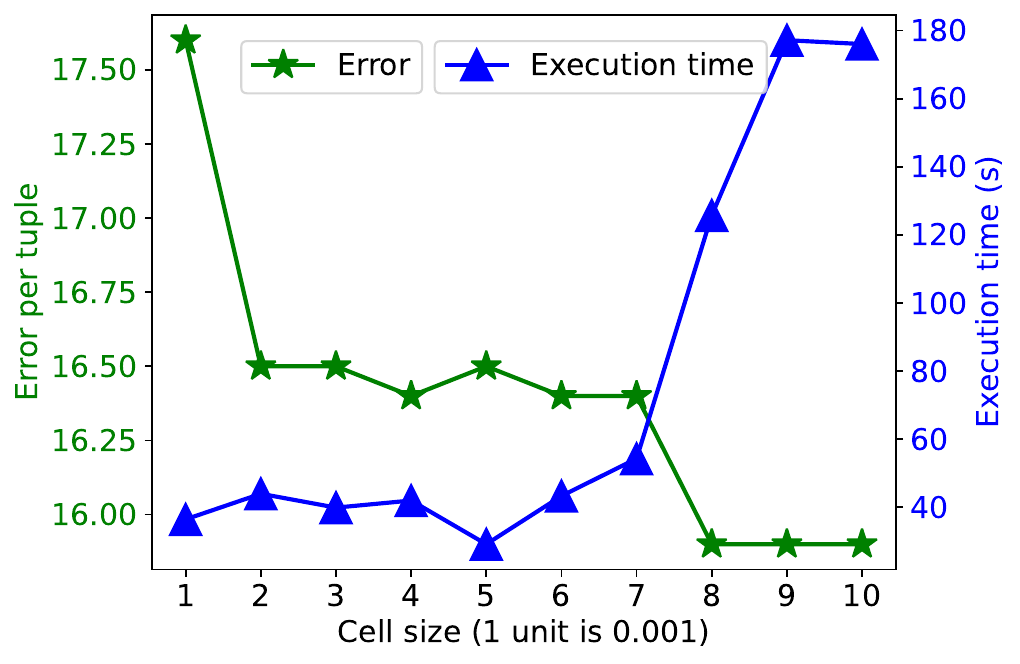}
        \caption{\NBA data, tradeoff of cell size}
        \label{fig:experiment_cell}
    \end{subfigure}

    \begin{subfigure}[h]{0.29\textwidth}
        \centering
        \includegraphics[width=1.05\textwidth]{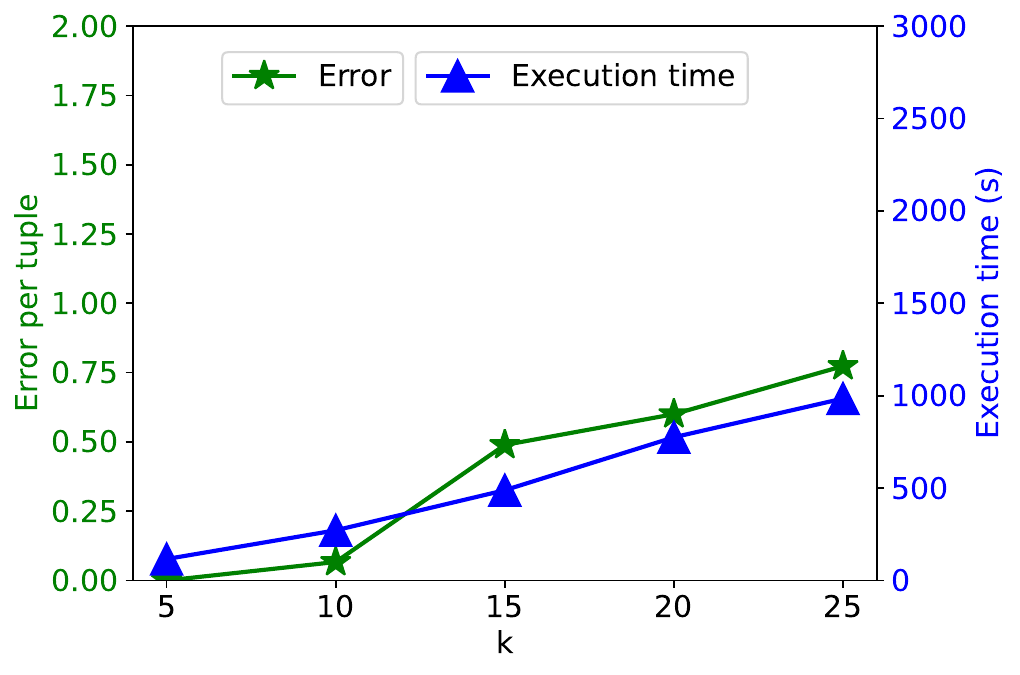}
        \caption{Uniform data}
        \label{fig:scalability_uniform}
    \end{subfigure}
    ~
    \begin{subfigure}[h]{0.29\textwidth}
        \centering
        \includegraphics[width=1.05\textwidth]{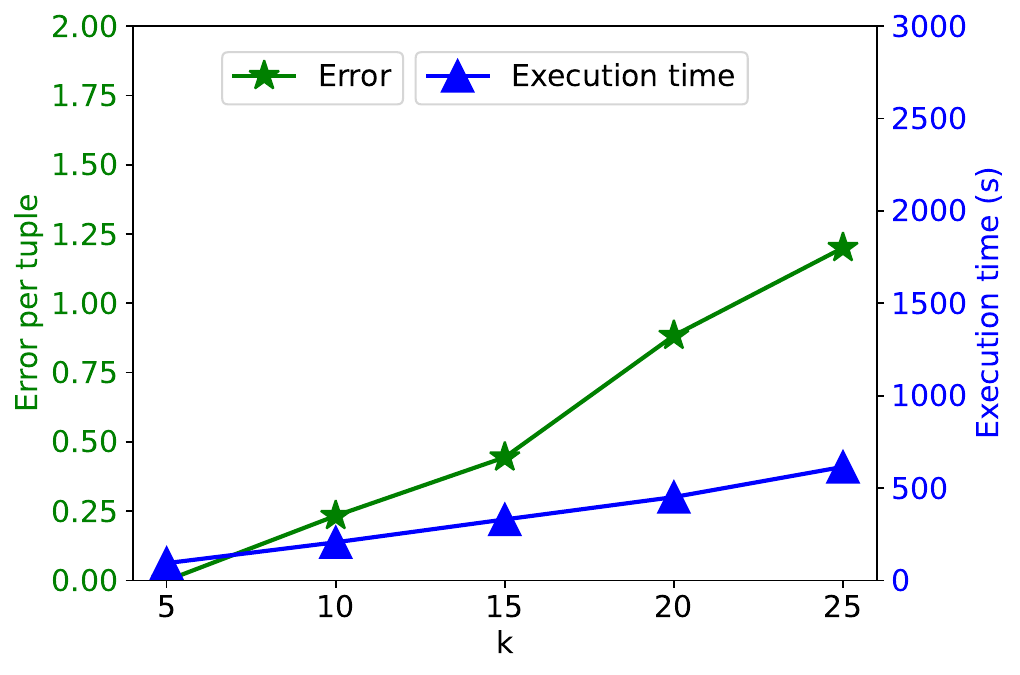}
        \caption{Correlated data}
        \label{fig:scalability_corrlated}
    \end{subfigure}
    ~
    \begin{subfigure}[h]{0.29\textwidth}
        \centering
        \includegraphics[width=1.05\textwidth]{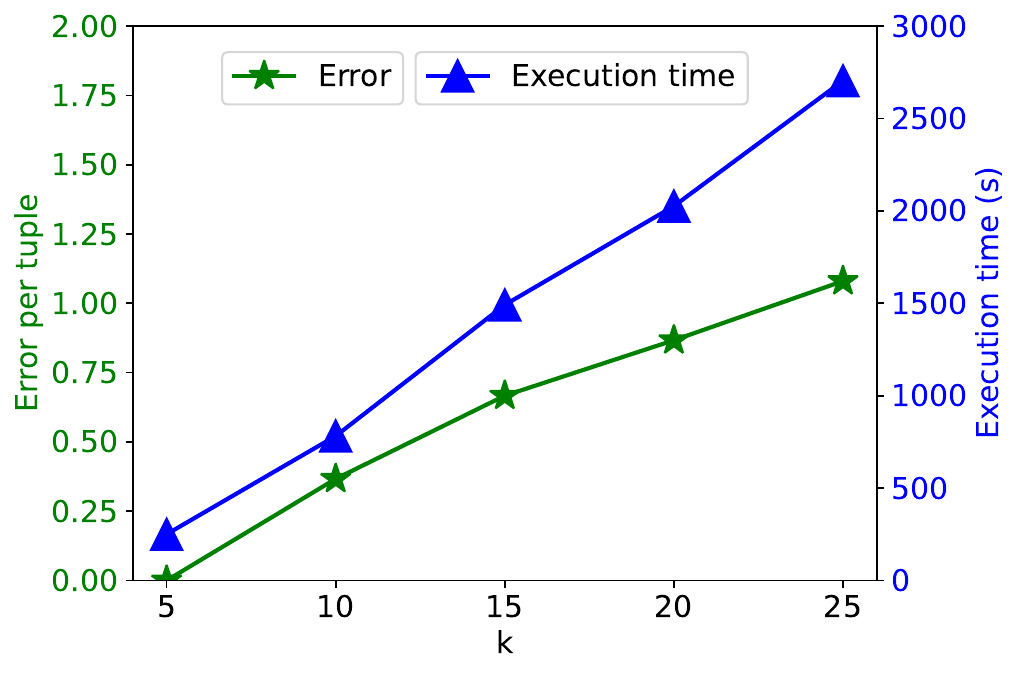}
        \caption{Anti-correlated data}
        \label{fig:scalability_anti_correlated}
    \end{subfigure}

    \begin{subfigure}[h]{0.29\textwidth}
        \centering
        \includegraphics[width=1.05\textwidth]{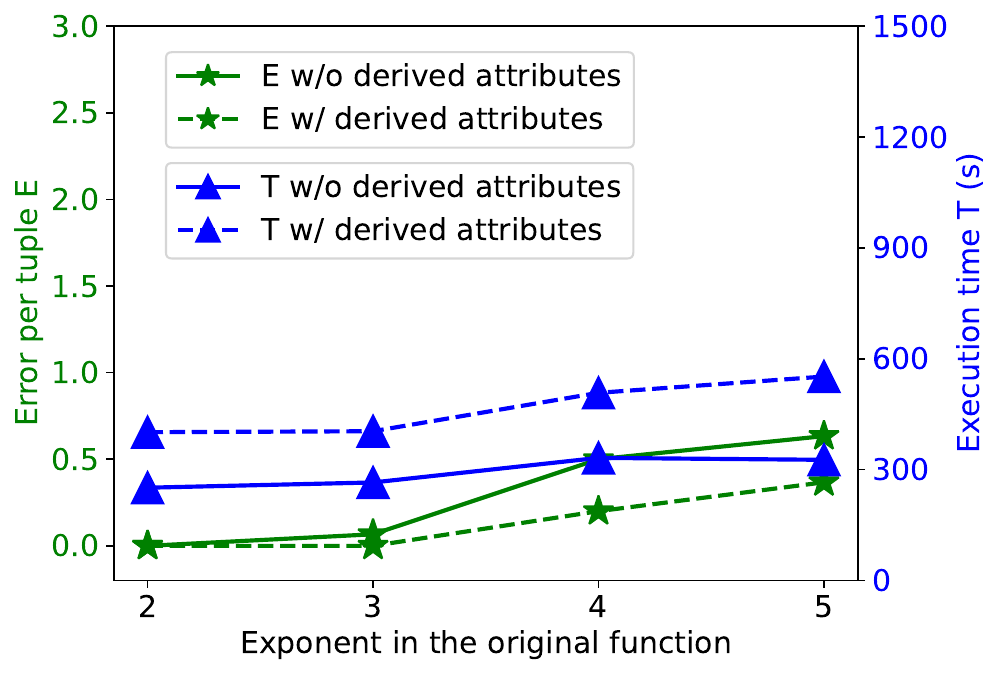}
        \caption{Uniform data}
        \label{fig:function_uniform}
    \end{subfigure}
    ~
    \begin{subfigure}[h]{0.29\textwidth}
        \centering
        \includegraphics[width=1.05\textwidth]{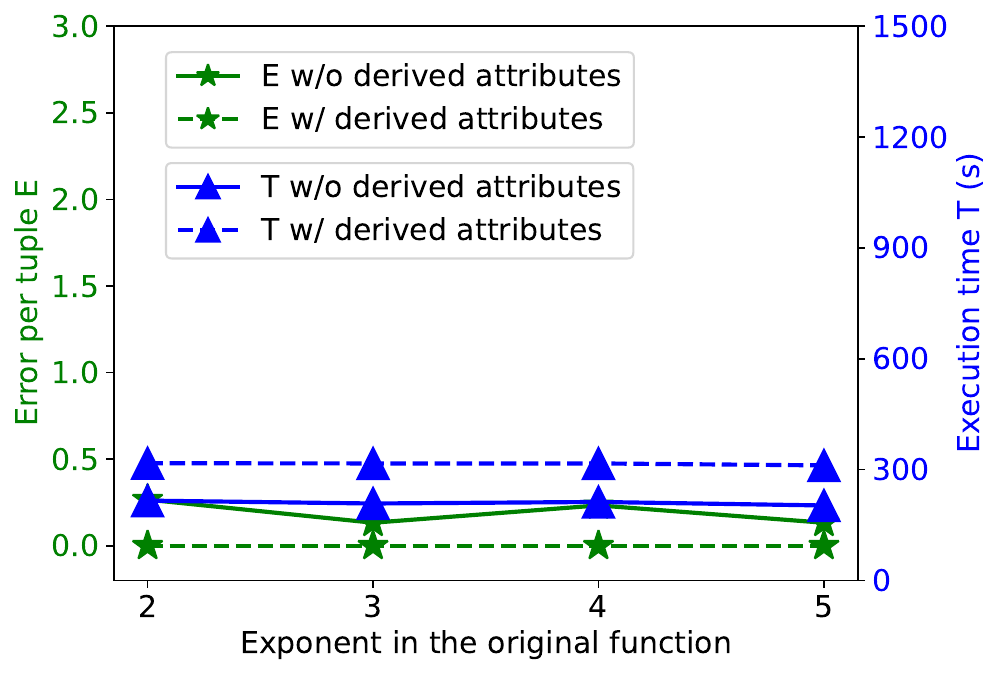}
        \caption{Correlated data}
        \label{fig:function_correlated}
    \end{subfigure}
    ~
    \begin{subfigure}[h]{0.29\textwidth}
        \centering
        \includegraphics[width=1.05\textwidth]{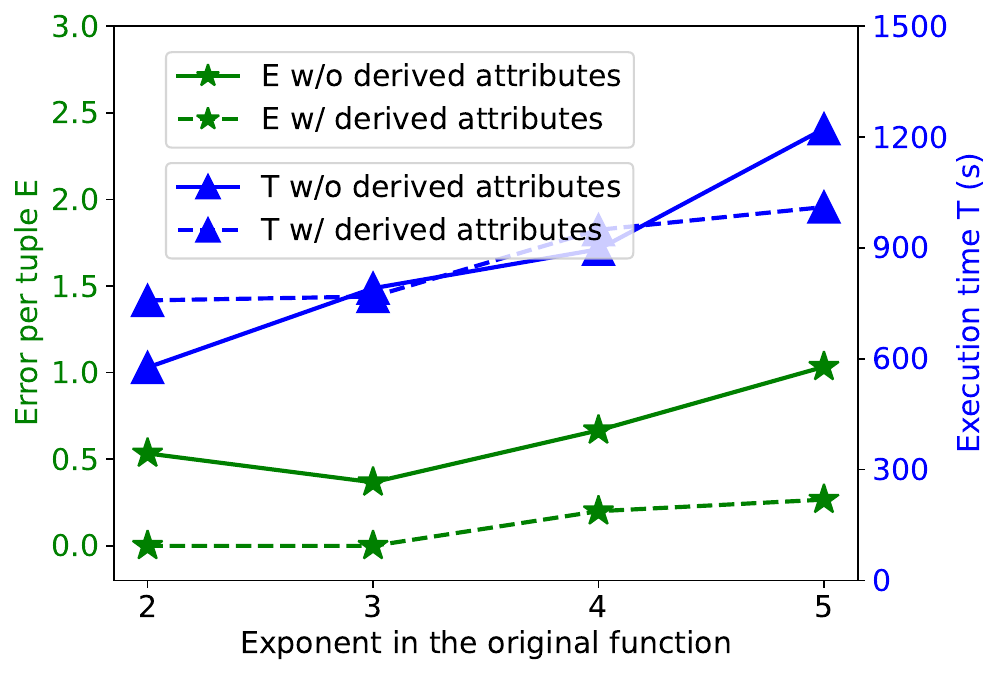}
        \caption{Anti-correlated data}
        \label{fig:function_anti-correlated}
    \end{subfigure}
\caption{Performance on \OPT}
\label{fig:experiments}
\end{figure*}

We compare the position error of our exact solution \RankHow against competitors
\Sampling, \LinearRegression, \OrdinalRegression and \Adarank on the \NBA (ranked by MP*PER)
and CSRankings datasets.
We present the big picture first, followed by detailed experimental results for
varying parameter setups.
To make \Sampling comparable to the other methods, we set a time budget equal to the time taken by \RankHow.
On the \NBA data, \Adarank always has at least twice larger error than the other approaches
including \LinearRegression, which it is supposed to improve by optimizing for the right
objective (position-based error instead of score-based error).
The reason is that one of the ranking attributes is much more
correlated with the given ranking than the others, resulting in that same attribute being
selected repeatedly as the weak ranker in each boosting round.
Hence for readability, we omit the \Adarank results from the plots for the \NBA data.

\introparagraph{Big picture (\Cref{fig:experiment_approaches})}
First a big picture of the performance of each approach on the \OPT problem is shown.
We set $m$ = 5, $k$ = 6 and use the entire \NBA dataset.
The figure shows that although \OrdinalRegression, \LinearRegression
and \Adarank are fast, their errors are far from the
minimal error returned by \RankHow.
The errors for \LinearRegression and \Adarank are actually 10.17 and 30,
but we placed those two points at the corner of the figure for better presentation.
Given more execution time, \Sampling tends to find solutions with lower error,
but it remains far from the minimum.
In addition to the exact solution, our approximate approach
\SymGD (\Cref{alg:sgd_cell} is used here for controlling the execution time) can generate the optimal result much faster.
The three points for \SymGD take 5, 11 and 15 seconds respectively.

\introparagraph{Varying $k$ (\Cref{fig:experiment_nba_k,fig:experiment_csrankings_k})}
For a larger $k$, average per-tuple error of all approaches increases.
This is expected, as it becomes harder for a linear function to represent
the ordering of more points accurately.
\RankHow performs better than all competitors for a larger $k$, and its errors on both
datasets are small (smaller than 4 positions per tuple) relative to the large data size.

\introparagraph{Varying $n$ (\Cref{fig:experiment_nba_n,fig:experiment_csrankings_n})}
As the number of tuples increases, the error remains stable, because adding
more lower-ranked tuples does not influence the top-$k$ tuples too much:
While the top-$k$ tuples should be in the correct position, for the
lower-ranked ones it suffices to rank them anywhere below the top-$k$.
\LinearRegression cannot exploit the weaker ordering requirements for
lower-ranked tuples, resulting in relatively greater impact of those tuples
on the solution, and thus steeper increase in error.

\introparagraph{Varying $m$ (\Cref{fig:experiment_nba_m,fig:experiment_csrankings_m})}
More attributes provide more choices to construct a more accurate
scoring function, therefore error should generally decrease with $m$.
\RankHow, being the method producing the optimal error, even
\emph{guarantees} that adding ranking attributes will never
increase error. 
\RankHow again dominates the competitors and even finds the perfect rankings when the number of attributes is sufficiently large.
Interestingly, \Sampling's error shows no clear trend, because more
attributes lead to a higher-dimensional weight space to sample from.

\subsection{Dealing with Numerical Issues}
\label{sec:exp_numerical}
We show \textbf{our approach for dealing with numerical issues is effective and general}.
Recall that \RankHow addresses numerical problems through thresholds
$\epsilon_1, \epsilon_2$.
In all experiments, we verify the solutions returned by the solver. 
All of them successfully passed
the verification, which shows the \emph{effectiveness} of our approach
for dealing with numerical imprecision.

We now show that numerical issues arise in practice,
if $\epsilon_1$ is set too low. This happens even for small data
as we demonstrate for a subset of the \NBA data, consisting of 10 tuples
and 8 attributes; as we vary $k$ from 1 to 10.
Following the approach discussed in \Cref{sec:numerical_issues}, we
set $\epsilon_1 = 10^{-4}, \epsilon_2 = 0$.
This is compared to a setting that essentially ignores numerical
issues, using $\epsilon_1 = 10^{-10}, \epsilon_2 = 0$.
\cref{tab:numerical} shows the results, reporting
the true position error as determined by the verification process.
For the larger $\epsilon_1$, \RankHow always returns the
perfect ranking, while the smaller $\epsilon_1$ causes ranking errors.
Essentially the solver believes to have found the optimal solution
(with zero error), but due to numerical problems, the scoring function
found actually produces a ranking that disagrees with the solver's
view of it.
To demonstrate the \emph{generality} of our numerical-correction approach,
we also apply it to \OrdinalRegression. The results mirror those for \RankHow:
When using a larger $\epsilon_1 = 10^{-4}$, \OrdinalRegression avoids picking
an inconsistent solution, which caused a position error for the smaller $\epsilon_1$.

\begin{table}[tb]
    \centering
    \caption{Position error for \RankHow and
    \OrdinalRegression (\OrdinalRegressionWith, \OrdinalRegressionWithout).
    The ``+'' version of the technique uses a sufficiently large $\epsilon_1 = 10^{-4}$
    to address numerical imprecision, while the ``-'' version
    works with a smaller $\epsilon_1 = 10^{-10}$. 
    }
    \begin{tabular}{|c|c|c|c|c|c|c|c|c|c|c|}
    \hline
    k & 1 & 2 & 3 & 4 & 5 & 6 & 7 & 8 & 9 & 10 \\
    \hline
    \RankHowWith & 0 & 0 & 0 & 0 & 0 & 0 & 0 & 0 & 0 & 0 \\
    \RankHowWithout & 0 & 1 & 1 & 3 & 0 & 1 & 2 & 4 & 2 & 0 \\
    \OrdinalRegressionWith & 0 & 0 & 0 & 0 & 0 & 0 & 0 & 0 & 0 & 0 \\
    \OrdinalRegressionWithout & 0 & 0 & 0 & 1 & 0 & 1 & 1 & 1 & 1 & 1 \\
    \hline
    \end{tabular}
    \label{tab:numerical}
\end{table}

\subsection{Approximation}
\label{sec:approximation}

We study running time and error when solving
\OPT approximately using \SymGD, which scales to larger $k$ and $n$.

\introparagraph{Approximation quality (\Cref{fig:experiment_approximation})}
To demonstrate the viability of \SymGD,
we show that conducting the symbolic gradient descent starting with a good seed cell
can be much faster than running the ``global'' \RankHow on the entire
space of weight vectors.
We ran \SymGD on the \NBA dataset for all experiments reported
in \Cref{fig:experiment_nba_k,fig:experiment_csrankings_n,fig:experiment_csrankings_m}.
In all cases we use \Cref{alg:sgd} with a fixed large cell size 0.1 to show both the execution time and error.
\Cref{fig:experiment_approximation} reports reduction in execution time vs increase
in per-tuple position error when using \SymGD vs the global version of \RankHow.
The ideal spot in the plot is the lower left corner, and the majority
of points is there. This means that seeding plus optimal local search
can find a scoring function with optimal or almost optimal error
in a fraction of the time of the global version---often by an order of magnitude or more.

\introparagraph{Impact of cell size (\Cref{fig:experiment_cell})}
We explore the tradeoff between approximation quality and cell size $c$
for $m = 8, k = 10$ on the entire \NBA data.
\Cref{fig:experiment_cell} shows that as cell size increases,
the error drops with little impact on execution time, until cell size
reaches 0.008 (8 units).
This representative result demonstrates that by adjusting cell size,
we can control the tradeoff between running time and result quality.

\subsection{Scalability and Generalizability}
\label{sec:scalability}

We use large synthetic datasets (1 million tuples each)
of three different distributions to
demonstrate the scalability of \SymGD and show that its error improves with additional derived attributes.
On all datasets, we set the cell size to 0.01 and $\epsilon_1 = 10^{-5}, \epsilon_2 = 0$.
For each distribution, reported results are the average over the three
datasets of the same distribution.

\introparagraph{Scalability (\Cref{fig:scalability_uniform,fig:scalability_corrlated,fig:scalability_anti_correlated})}
All datasets are sorted by a nonlinear ranking function ${\sum _i^m{A_i ^{3}}}$.
Despite using a simpler linear function,
\SymGD performs well, even for large $k$ taking less than 1 hour and delivering
a low error of no more than 1.5 positions per tuple on average.

\introparagraph{Generalizability (\Cref{fig:function_uniform,fig:function_correlated,fig:function_anti-correlated})}
Different nonlinear ranking functions from ${\sum _i^m{A_i ^{2}}}$ to
${\sum _i^m{A_i ^{5}}}$ are used to generate the given ranking for
datasets of all three distributions.
The results show that with only the five original attributes (solid lines),
a linear ranking function with low error of at most 1.1 positions per tuple
can always be found.
To further reduce the error of \SymGD and show the good generalizability
of linear ranking functions, we added five derived attributes $A_i^2, i=1,\ldots, 5$,
to relation $R$. With the derived attributes, we see much lower error
with moderatly higher execution time.
On correlated data, with the derived attributes, \SymGD is able to find
a perfect \emph{linear} ranking function for rankings generated by all
nonlinear ranking functions, even when the true ranking functiong
had a higher exponent like 5.

\section{Related Work}\label{sec:related}

\introparagraph{Ordinal regression}
The approach most closely related to \RankHow is the ordinal regression
method of Srinivasan~\cite{76-jacm-srinivasan}. It also relies on a linear program (without integer variables) to identify a linear scoring
function for a given ranking. The program does not allow ties and
does not address numerical issues.
Furthermore, it does not model tuple ranks through indicator variables, but instead
relies on real-valued slack variables that keep track of the score difference for all tuple pairs
that are inverted. Hence the program optimizes for a score-based loss,
not position-based error.
This provides only weak guarantees about the
quality of the resulting ranking. Consider a scenario where the top-100
tuples of the given ranking receive very similar scores, such that the
score difference between any pair of them is at most 0.0001.
Then a score assignment that moves all top-10 tuples down 90 positions
would result in a total score penalty below 0.1. The same penalty is
incurred when swapping two adjacent tuples whose scores differ by 0.1.
The corresponding position-based errors are 1800 and 2,
respectively, capturing properly the impact of moving 10 tuples down by
90 positions each vs having only 2 tuples swap positions.

\introparagraph{Ordinal classification}
Recent work on ordinal ``regression'' can be more accurately described
as ordinal \emph{classification}~\cite{16-tkde-gutierrez} and is similar to pointwise learning-to-rank.
The common goal is to learn a function
and threshold values, so that all function values falling into the
$i$-th range are classified as the $i$-th highest value of an ordinal
response variable. Ordinal classification approaches
cannot be applied in a meaningful way to our problem, because they
expect \emph{many tuples per value of the ordinal response variable for training}.
Intuitively, they are designed for problems like predicting a person's
product preference on a scale from ``very low'' to ``very high'',
based on previous purchases and product ratings.

\introparagraph{Explainable learning-to-rank (LtR)}
In the LtR literature, many different notions of ``explanation''
exist~\cite{22-arxiv-anand}, including various types of
attribute-importance measures, text snippets that explain
why a document is relevant for a search query, and simple scoring
functions. The latter can produce linear functions
\cite{19-wsdm-singh,21-wsdm-zhuang,16-kdd-ribeiro}
and thus are most related to our work.
In contrast to \RankHow (\Cref{eq:best_LP_topk}) those methods are
heuristics that do not guarantee optimality.
And like practically any machine-learning method for ranking or prediction,
they do not support constraints for exploring alternative scoring functions.
Note that competitors \LinearRegression and \Adarank in our experiments cover the spectrum
of explainable LtR techniques when applied to our problem.
As our experiments demonstrate, \RankHow beats them.
To see why \LinearRegression and \Adarank represent the
explainable LtR methods,
first consider post-hoc explainable LtR: Given a complex accurate
ranking model, they analyze or probe that model to generate a
simple explanation for it~\cite{22-arxiv-anand}.
As discussed in \Cref{sec:introduction}, a perfect complex model
for our ranking problem is the given ranked data relation
$R$ itself. When applying a post-hoc explainable LtR approach to it,
this is equivalent to learning the linear scoring function from the data.
In contrast to post-hoc approaches, an interpretable-by-design (IBD) method \cite{21-wsdm-zhuang}
directly learns a generalized additive model that does not involve any interaction
between features to maintain interpretability.
When each sub-function on an attribute is in its simplest form---a linear function---the
model also learns a linear scoring function from the data.
Hence, the most meaningful application
of these explainable LtR methods in our context is to apply
\LinearRegression or \Adarank to the ranked relation $R$.

\introparagraph{Ranking explanation}
Why-not-yet \cite{23-pvldb-chen} identifies linear functions that
place a given tuple into the top-$k$.
He, Lo et al. explore why-not problems for top-$k$ queries~\cite{12-icde-he,14-tkde-he,16-tkde-xu},
which return a refined query that includes the missing tuple in the top-$k$, but allows
the use of a larger $k$ in addition to modifying the ranking function.
This is different from \OPT, because we are interested in approximating an
entire ranking, not reorder its top-$k$ tuples.
Vlachou et al.~\cite{10-icde-vlachou,11-tkde-vlachou} introduce the problem of 
reverse top-$k$ queries. The monochromatic version looks for
\emph{all} $W$ that rank a query tuple in the top-$k$.
The authors only provide an exact solution for $m = 2$ and acknowledge the hardness of finding an exact solution in higher dimensions \cite{11-tkde-vlachou}.
In the bichromatic version, the weight vectors must be chosen
from a given set of candidates.
Similar to the monochromatic reverse top-$k$ problem,
Asudeh et al. \cite{19-sigmod-asudeh-fair-ranking, 18-vldb-asudeh-stable-rankings} introduce
an approach for identifying \emph{all} satisfiable regions by cutting the space with
hyperplanes. We show in our experiments that using an MILP solver achieves orders of magnitude
running time improvement compared to their PTIME  algorithm
(extended for solving \OPT).
Wang et al.~\cite{24-icde-wang} study the similar problem of
reverse regret query, which identifies all tuples that ``barely'' missed
the top-$k$ based on a linear scoring function.
Panev et al. \cite{16-edbt-panev,16-pvldb-panev} reverse-engineer top-$k$ queries
by applying different selection conditions and aggregation functions,
but do not support linear combinations of multiple attributes.
{Many approaches for determining the importance of attributes for
a ranking have been proposed in the explainable LtR~\cite{19-sigir-verma,22-arxiv-anand}
and database community~\cite{24-arxiv-pliatsika,20-pvldb-gale}.
They are interesting in their own right, but do not provide a scoring
function. In fact, \cite{24-arxiv-pliatsika} needs a scoring function
to compute Shapley values.

\introparagraph{Exact MILP solvers}
Recent work on numerically exact MILP
solvers~\cite{eifler2023computational,cook2013hybrid}
could address the numerical issues described in \Cref{sec:numerical_issues}.
Unfortunately, the current state-of-the-art tools like SCIP~\cite{scip}
are not yet capable of providing performance
competitive with MILPs that exclusively use floating point computation.

\introparagraph{Synthesis of linear ranking functions}
Work with similar-sounding terminology \cite{01-tacas-colon,04-vmcai-podelski,12-infcomput-bagnara}
explores \emph{program termination problems}: A ``ranking'' function
assigns decreasing values to state transitions caused by loop
execution, proving termination when those values are lower-bounded.
The function is defined by a system of inequalities that
are extracted from the loop code, hence not related to our ranking problem.

\section{Conclusion}

We formally define the \OPT problem of synthesizing linear scoring functions
with flexible constraints on their weights, and propose \RankHow to
solve it exactly. Our MILP-based solution, despite the theoretical
NP-hardness of general MILP, is orders of magnitude faster than
a PTIME implementation that reduces the problem to a series of LPs.
This makes a strong case for revisiting previous
work~\cite{19-sigmod-asudeh-fair-ranking,24-icde-wang} that
focused on techniques similar to our PTIME construction.
To overcome numerical issues, which can severely affect
a ranking, we introduce ``gap'' parameters 
in \Cref{eq:best_LP_topk}. Our experiments show that other techniques
relying on (MI)LP solvers for ranking problems would also benefit
from this construction.
To improve scalability, we propose \SymGD. It is a novel gradient-descent
paradigm that exploits the structure of \OPT to speed up search for local optima,
thus offering a way to control the tradeoff between running time and result quality.

\section*{Acknowledgment}

This material is based upon work supported by the National Science Foundation under
Grant No.~1956096.

\bibliographystyle{IEEEtran}
\IEEEtriggeratref{59}
\bibliography{sample}

\end{document}

%% file: datalabmacros.tex
\usepackage[a-2b]{pdfx}   

\usepackage{amsmath}
\usepackage{amsthm}
\usepackage{mathtools}
\usepackage{graphicx} 
\usepackage{url}
\usepackage{enumitem} 

\usepackage{stmaryrd} 

\usepackage{upgreek} 
\usepackage{bbding} 

\usepackage{listings} 
\usepackage{tabularx} 
\usepackage{booktabs} 
\usepackage{multirow}
\usepackage{hhline} 
\usepackage{diagbox}
\usepackage{pgfplotstable} 

\usepackage{xcolor,colortbl}    

\usepackage{tikz} 
\usetikzlibrary{matrix}
\usetikzlibrary{calc}
\usetikzlibrary{math}
\usetikzlibrary{arrows.meta,positioning}
\usetikzlibrary{intersections, pgfplots.fillbetween}

\usepackage{easybmat} 

\usepackage{adjustbox}
\def\eox{\unskip\kern 10pt{\unitlength1pt\linethickness{.4pt}$\diamondsuit${}}} 

\usepackage{rotating}		

\usepackage{xspace} 

\usepackage{subcaption} 
\hypersetup{                                                            
	pdfpagemode=UseNone,               
    pdfstartview=Fit,
    pdfpagelayout=SinglePage,       
    bookmarks=false,
    bookmarksnumbered=true,
    bookmarksopen=false,             
    pdftex=true,
    unicode,
    colorlinks=true,       
    linkcolor=blue,          
    citecolor=cyan,  
    filecolor=magenta,      
     urlcolor=blue           
     plainpages=false,           
     hypertexnames=true,         
     pdfpagelabels=true,         
     hyperindex=true,
     naturalnames=false     
}

\usepackage[capitalise,nameinlink]{cleveref} 

\usepackage{aliascnt}  	

\newtheorem{example}{Example}
\newtheorem{definition}{Definition}
\newtheorem{lemma}{Lemma}
\newtheorem{theorem}{Theorem}

\AtEndPreamble{%
    \hypersetup{colorlinks,
      linkcolor=purple,
      citecolor=blue,
      urlcolor=teal}}

\usepackage{tcolorbox}		
\tcbuselibrary{breakable,skins}		

\tcbset{examplestyle/.style={
		enhanced jigsaw,	
		colback=blue!10,	
		colframe=blue!10,	
		arc=0mm,
		boxrule=0pt,		
		left=1mm,
		right=1mm,
		left skip= 0mm,  
		right skip= 0mm, 
		top=-2pt,		
		bottom=2pt,		
		breakable,		
		parbox = false,		
		before={\par\pagebreak[0]\vspace{1mm}\parindent=0pt},		
		after={\par\pagebreak[0]\vspace{1mm}\parindent=0pt},				
		bottomrule = 0mm,
		boxsep = 0mm,					
		topsep at break=0pt,			
		bottomsep at break=0pt,			
		pad at break=0mm,
		pad before break=1mm,		
		pad after break=1mm,		
		bottomrule at break=0mm,
		toprule at break=0mm,		
		}}

\usepackage{footnote}

\tcolorboxenvironment{example}{examplestyle}

\makeatletter
\DeclareRobustCommand*\uell{\mathpalette\@uell\relax}
\newcommand*\@uell[2]{
  \setbox0=\hbox{$#1\ell$}
  \setbox1=\hbox{\rotatebox{10}{$#1\ell$}}
  \dimen0=\wd0 \advance\dimen0 by -\wd1 \divide\dimen0 by 2
  \mathord{\lower 0.1ex \hbox{\kern\dimen0\unhbox1\kern\dimen0}}
}
\makeatother

\usepackage{marginnote}

\newcommand{\introparagraph}[1]{\textbf{#1.}} 

\renewcommand{\epsilon}{\varepsilon} 

\renewcommand{\O}{{\mathcal{O}}} 

\usepackage{scalerel}
\usepackage{tikz}
\usetikzlibrary{svg.path}

\definecolor{orcidlogocol}{HTML}{A6CE39}
\tikzset{
  orcidlogo/.pic={
    \fill[orcidlogocol] svg{M256,128c0,70.7-57.3,128-128,128C57.3,256,0,198.7,0,128C0,57.3,57.3,0,128,0C198.7,0,256,57.3,256,128z};
    \fill[white] svg{M86.3,186.2H70.9V79.1h15.4v48.4V186.2z}
                 svg{M108.9,79.1h41.6c39.6,0,57,28.3,57,53.6c0,27.5-21.5,53.6-56.8,53.6h-41.8V79.1z M124.3,172.4h24.5c34.9,0,42.9-26.5,42.9-39.7c0-21.5-13.7-39.7-43.7-39.7h-23.7V172.4z}
                 svg{M88.7,56.8c0,5.5-4.5,10.1-10.1,10.1c-5.6,0-10.1-4.6-10.1-10.1c0-5.6,4.5-10.1,10.1-10.1C84.2,46.7,88.7,51.3,88.7,56.8z};
  }
}

\RequirePackage{etoolbox}
\DeclareRobustCommand\orcidicon[1]{\href{https://orcid.org/#1}{\mbox{\scalerel*{
\begin{tikzpicture}[yscale=-1,transform shape]
\pic{orcidlogo};
\end{tikzpicture}
}{|}}}}

%% file: individualmacros.tex
\usepackage{array}
\usepackage{makecell}

\newcommand{\RankHow}{\textsc{RankHow}\xspace}
\newcommand{\SymGD}{\textsc{Sym-GD}\xspace}

\newcommand{\OPT}{\textsc{OPT}\xspace}

\newcommand{\Sampling}{\textsc{Sampling}\xspace}
\newcommand{\LinearRegression}{\textsc{Linear Regression}\xspace}
\newcommand{\OrdinalRegression}{\textsc{Ordinal Regression}\xspace}
\newcommand{\Adarank}{\textsc{AdaRank}\xspace}
\newcommand{\Tree}{\textsc{Tree}\xspace}

\newcommand{\NBA}{\textsc{NBA}\xspace}

\newcommand{\RankHowWith}{\textsc{RankHow+}\xspace}
\newcommand{\RankHowWithout}{\textsc{RankHow-}\xspace}
\newcommand{\OrdinalRegressionWith}{\textsc{OR+}\xspace}
\newcommand{\OrdinalRegressionWithout}{\textsc{OR-}\xspace}

\newcommand{\pred}{\ensuremath{\mathcal{P}}}
\newcommand{\rank}{\ensuremath{\rho}}
\newcommand{\score}{\ensuremath{f}}

\usepackage{algorithm}
\usepackage{orcidlink}